\documentclass[10pt]{article} 
\usepackage[affil-it]{authblk}
\usepackage[]{graphicx}
\usepackage{amsfonts}
\usepackage{fullpage}
\usepackage{amsmath}
\usepackage{natbib}

\title{Simultaneous reconstruction and displacement estimation for spectral-domain optical coherence elastography} 
\date{January 23 2019}
\author{Jonathan H. Mason\textsuperscript{1}}\author{Yvonne Reinwald\textsuperscript{2}}\author{Ying Yang\textsuperscript{3}}\author{Sarah Waters\textsuperscript{4}}\author{Alicia El Haj\textsuperscript{5}}\author{Pierre O. Bagnaninchi\textsuperscript{1}}
\affil{\textsuperscript{1}MRC Centre for Regenerative Medicine, The University of Edinburgh, 5 Little France Drive, Edinburgh, United Kingdom}
\affil{\textsuperscript{2}Department of Engineering, College of Science and Technology, Nottingham Trent University, Nottingham, United Kingdom}
\affil{\textsuperscript{3}Institute for Science and Technology in Medicine, Keele University, Keele, United Kingdom}
\affil{\textsuperscript{4}Mathematical Institute, University of Oxford, Oxford, United Kingdom}
\affil{\textsuperscript{5}Healthcare Technologies Institute, University of Birmingham, Birmingham, United Kingdom}

\begin{document} 
\maketitle 

\begin{abstract}
	Optical coherence elastography allows the characterization of the mechanical properties of tissues, and can be performed through estimating local displacement maps from subsequent acquisitions of a sample under different loads. This displacement estimation is limited by noise in the images, which can be high in dynamic systems due to the inability to perform long exposures or B-scan averaging. In this work, we propose a framework for simultaneously enhancing both the image quality and displacement map for elastography, by motion compensated denoising with the block-matching and 4D filtering (BM4D) method, followed by a re-estimation of displacement. We adopt the interferometric synthetic aperture microscopy (ISAM) method to enhance the lateral resolution away from the focal plane, and use sub-pixel cross correlation block matching for non-uniform deformation estimation. We validate this approach on data from a commercial spectral domain optical coherence tomography system, whereby we observe an enhancement of both image and displacement accuracy of up to 33\% over a standard approach.  
\end{abstract}

\section{INTRODUCTION}
Characterizing the mechanical properties of materials has many applications in industrial testing and bioengineering. These can be inferred by tracking the deformation of a specimen under known loads. Optical coherence tomography (OCT) is a powerful modality for the role of displacement tracking, due to its high resolution and non-invasive 3D imaging capability, where its use is known as optical coherence elastography (OCE)\cite{Kennedy2017}.

The key to performing OCE is calculating local displacements through a sequence of temporal images. This problem is actively treated in other image processing applications: known as `optical flow' in video compression; and `particle image velocimetry' in fluid dynamics. A common approach is block matching, whereby sub-images are matched through some metric, such as cross correlation.

The accuracy of block matching is limited by noise in the image sequence, which can be especially high in cases of rapid acquisitions. Additionally, if the displacement between was known to a high accuracy, then the noise could be reduced, in a manner similar to B-scan averaging. These concepts motivate a simultaneous approach to imaging, whereby denoising and motion estimation are combined.

An instance of this same concept was realised by Buades et al.\cite{Buades2016}, applied to video enhancement, where they performed patch based denoising on motion compensated frames. In a second step, they then recalculate the displacement on the cleaned frames, followed by another motion compensated denoising.

In this article, we propose to perform motion compensated denoising of OCT images by applying the BM4D method \cite{Maggioni2013} to warped frames from a temporal sequence. These enhanced images will then form the basis for a second motion estimation step. To enable good spatial resolution throughout the specimen, we will apply the ISAM method \cite{Ralston2007}, which we implement through the non-uniform fast Fourier transform (NUFFT) \cite{Fessler2003a}.  

\section{METHODOLOGY}
\subsection{Motion Compensated Denoising}\label{sec:method}
\begin{figure}
	\begin{center}
		\begin{tabular}{c}
			\includegraphics[height=7cm]{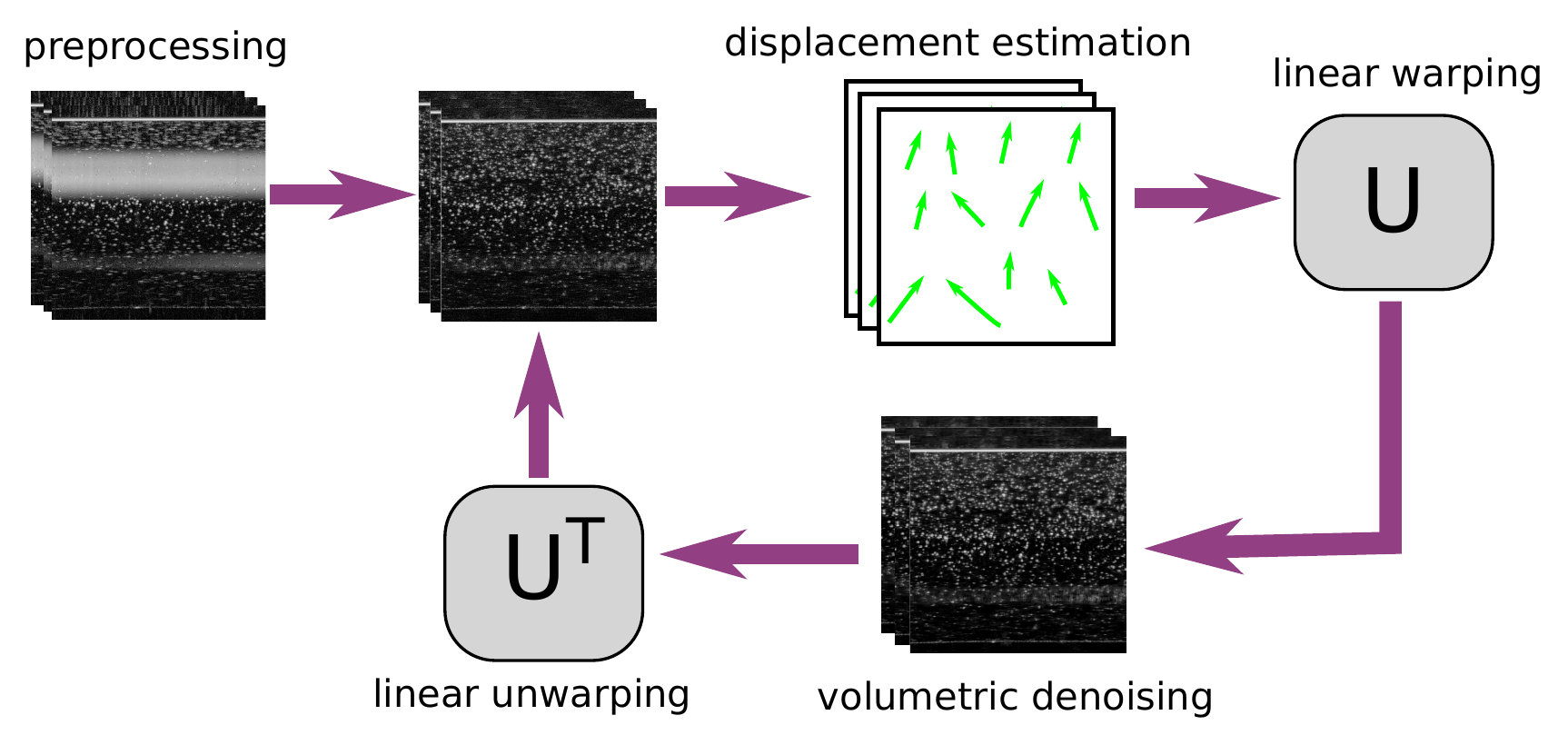}
		\end{tabular}
	\end{center}
	\caption[example] 
	{ \label{fig:flow_diagram} 
		Flow diagram of simultaneous displacement estimation and image enhancement.}
\end{figure}

A diagrammatic representation of the method is shown in Figure~\ref{fig:flow_diagram}. This shows the potential to iterate the displacement estimation and subsequent motion compensated denoising several times. In practice, we find that a single iteration, up to the re-estimation of the motion, is sufficient. In this case, the steps may be described as:
\begin{enumerate}
	\item Form initial images from standard reconstruction, including any appropriate preprocessing steps such as dispersion encoded artifact removal\cite{Hofer2009} or ISAM\cite{Ralston2007} resampling.
	\item Estimate relative displacement between neighboring frames with block matching, as a linear operation, $U$.
	\item Warp each neighboring frame to common spatial position through applying $U$.
	\item Perform motion compensated denoising, through the BM4D method \cite{Maggioni2013}.
	\item Apply adjoint of linear warping on denoised images, through $U^T$. 
	\item Recalculate displacement on denoised frames.
	\item Return to step 3 or terminate.
\end{enumerate}

When this process is repeated for several iterations, one should ensure that the linear warping operation, $U$, is applied to the original preprocessed images, and not the denoised frames. Otherwise, loss of information is likely.

\subsection{Linear Sub-pixel Displacement Estimation}
A key element of our framework is the means to calculate the local displacement throughout a sequence of images, from which we can derive a linear operator, $U$. For the displacement estimation, we use multi-resolution cross correlation block matching from the PIVlab\cite{Thielicke2014} Matlab toolbox. This software implements sub-pixel estimation through 2D Gaussian regression\cite{Nobach2005}. We then use linear interpolation to upsample from the block size resolution to the resolution of the image.

After the dense local deformations have been found, we form the operator, $U$, to bi-linearly interpolate a pixel's value to its new sub-pixel location. To preserve linearity, we have to ensure the displacements are constrained to the image view. With this, we do not apply any deformation to pixels that are not present in both images.

\section{EXPERIMENTATION}
\subsection{Displacement Estimation Validation} \label{sec:valid}
In this first experiment, we wish to validate our simultaneous flow estimation and image denoising. To this end, we induce known motion to a sample, and evaluate the accuracy in both image quality and local displacement.

Our sample in this experiment is a 2\% agarose gel with dispersed latex micro-beads, through which a 2 mm cross section was recorded. The specimen was placed on an motorized micrometer uniaxial stage, which was programmed to move laterally at steps of 10 $\mu$m at 5 positions. The imaging equipment used was a Wasatch photonics 800 nm OCT system.

\begin{figure}
	\begin{center}
		\begin{tabular}{cc}
			standard IFFT reconstruction & ISAM \\
			\includegraphics[height=6cm]{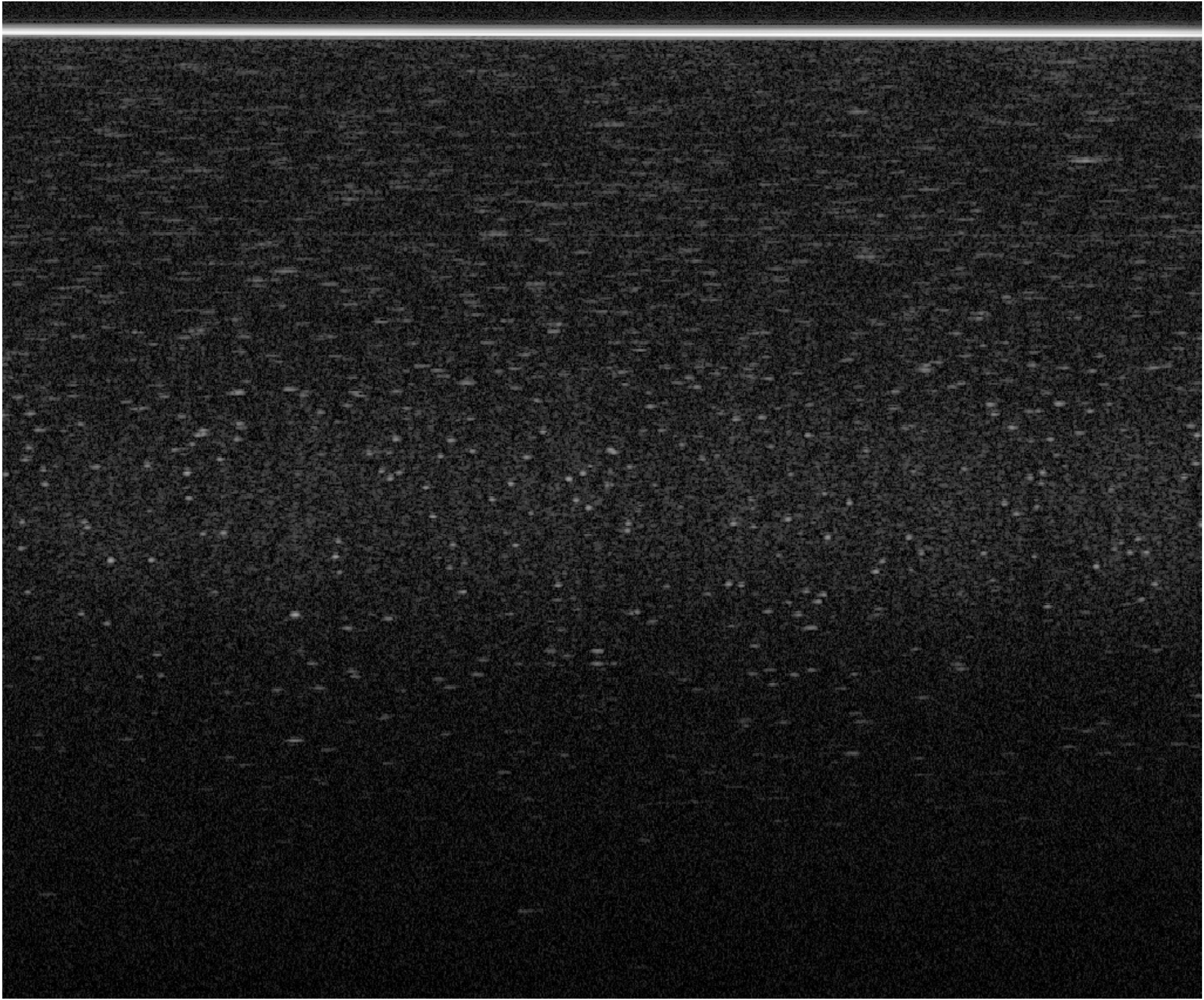}
			&\includegraphics[height=6cm]{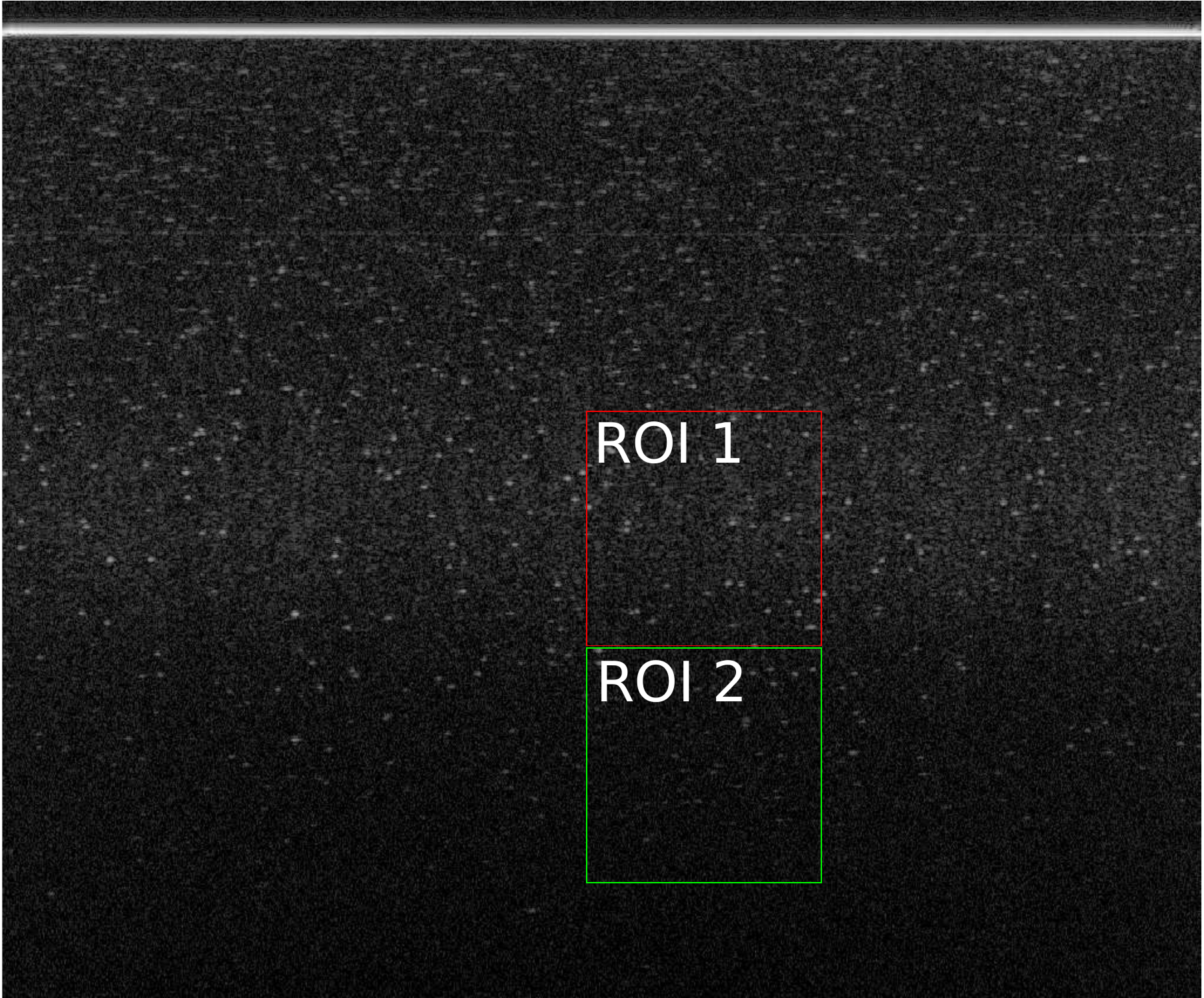}
		\end{tabular}
	\end{center}
	\caption[example] 
	{ \label{fig:sample_latex} Latex micro-bead specimen, and its ISAM reconstruction.}
\end{figure}

We adjusted the system, so that the focal point lay at the centre of the sample. This allowed the simple application of ISAM\cite{Ralston2007}, which has the effect of refocussing throughout the sample, and acts as our preprocessing operation. The original reconstruction of the sample, along with ISAM, are shown in Figure~\ref{fig:sample_latex}.

\begin{figure}[htb!]
	\begin{center}
		\begin{tabular}{c|c|c|c|c|}
			& single frame & 20 frame average & 5 frame warped average & 5 frame warped denoised \\
			\hline
			\rotatebox{90}{ROI 1}&
			\includegraphics[height=3.5cm]{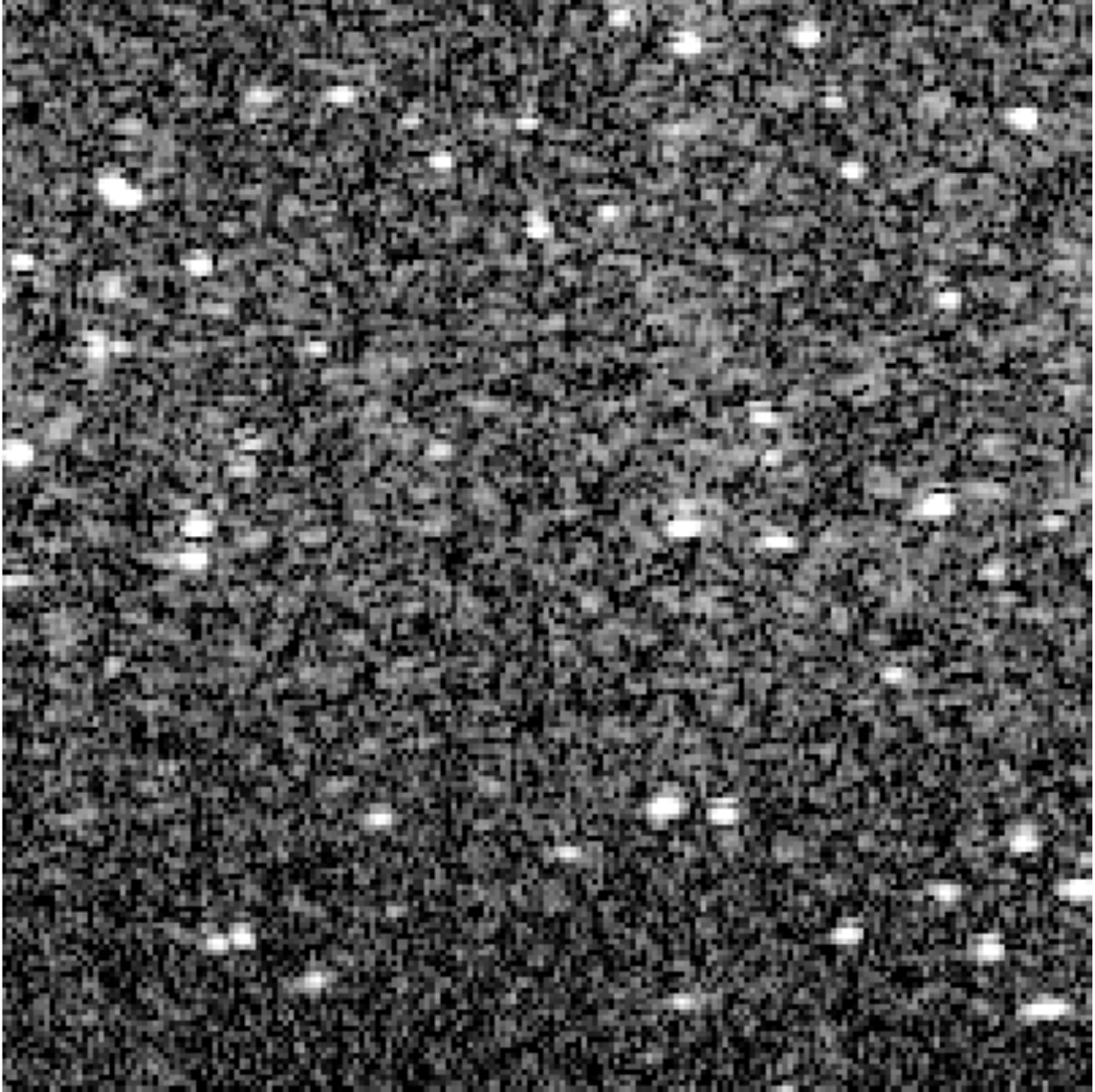}&
			\includegraphics[height=3.5cm]{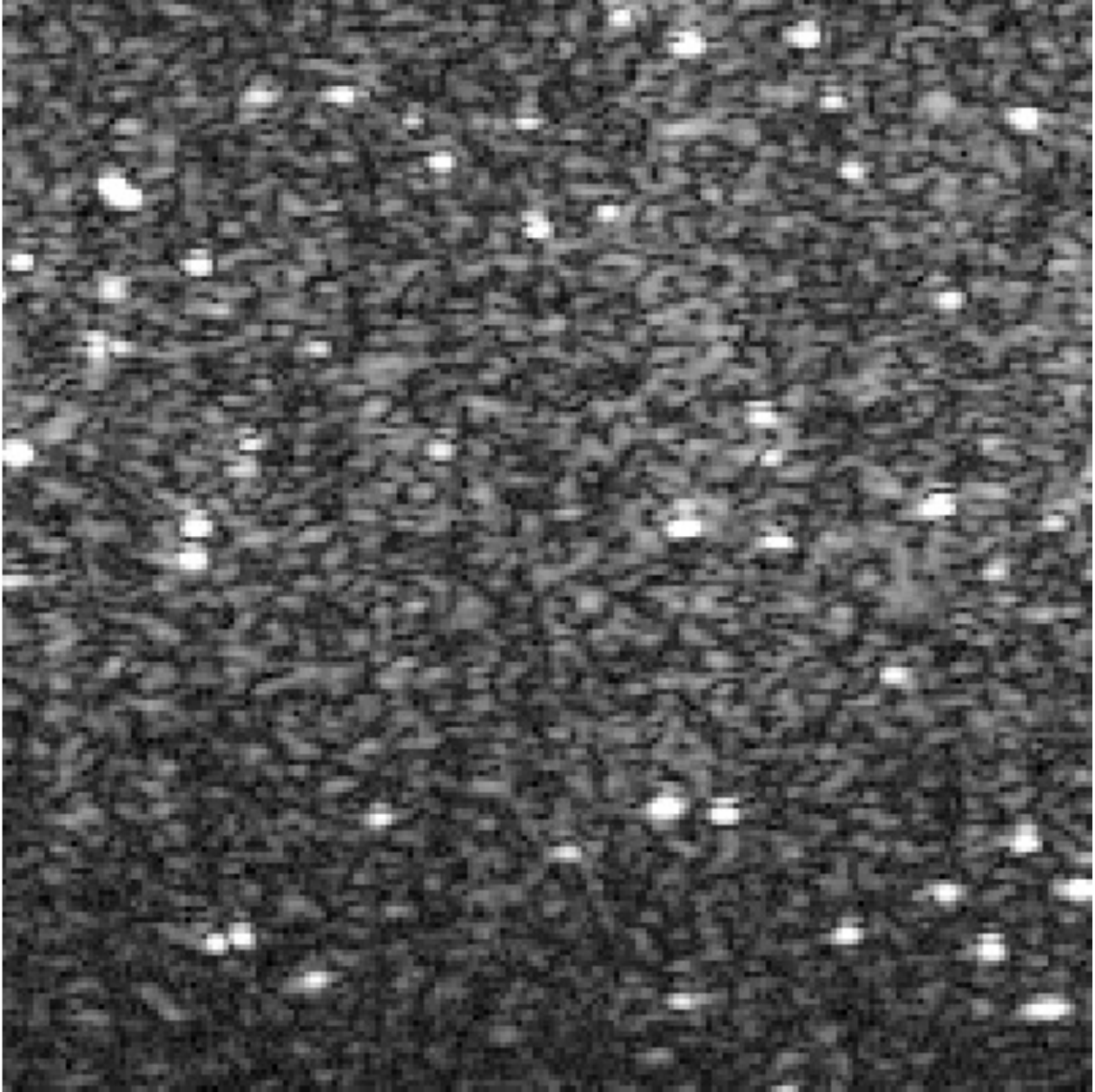}&
			\includegraphics[height=3.5cm]{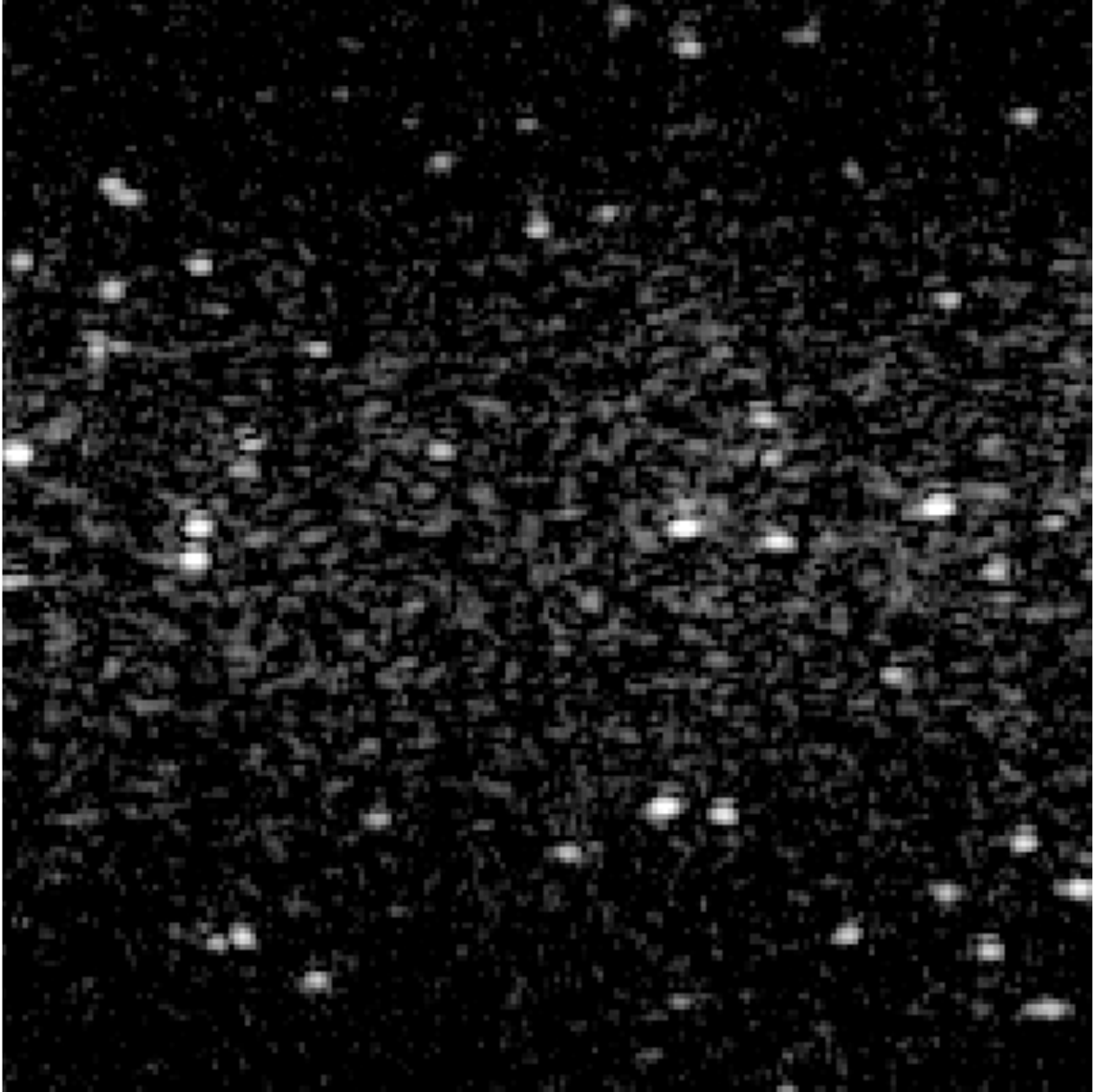}&
			\includegraphics[height=3.5cm]{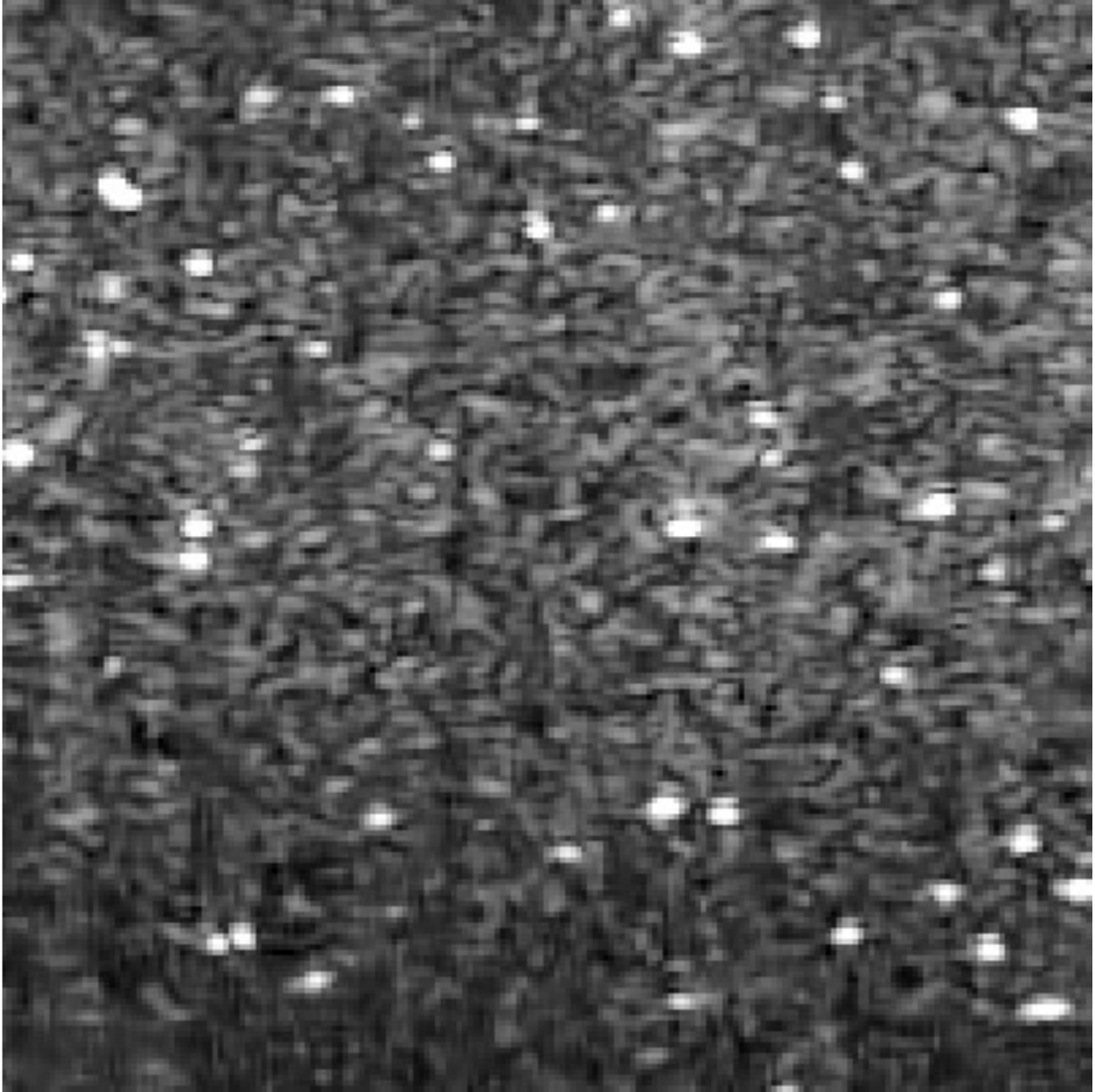}\\
			\hline
			\rotatebox{90}{ROI 2}&
			\includegraphics[height=3.5cm]{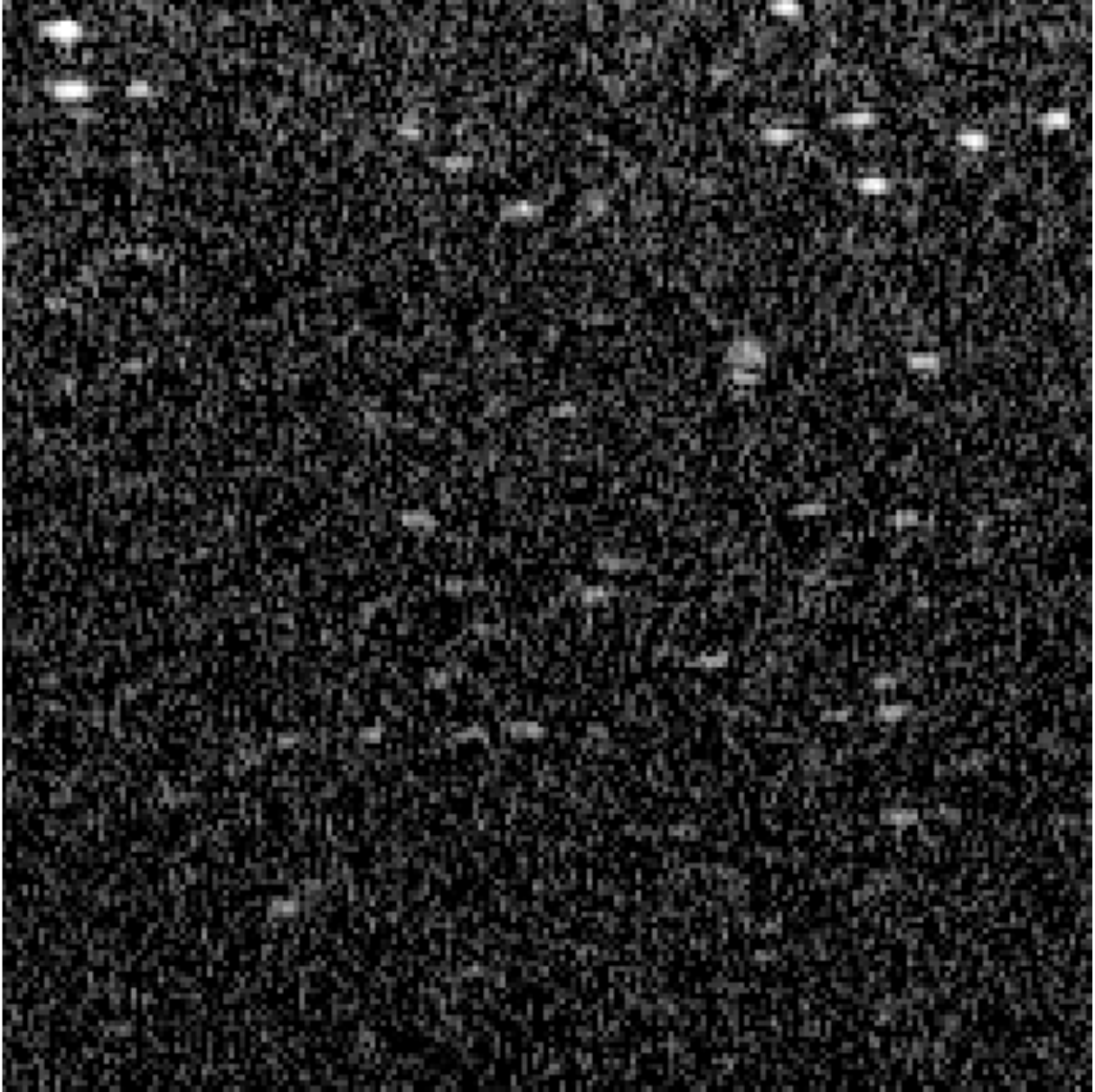}&
			\includegraphics[height=3.5cm]{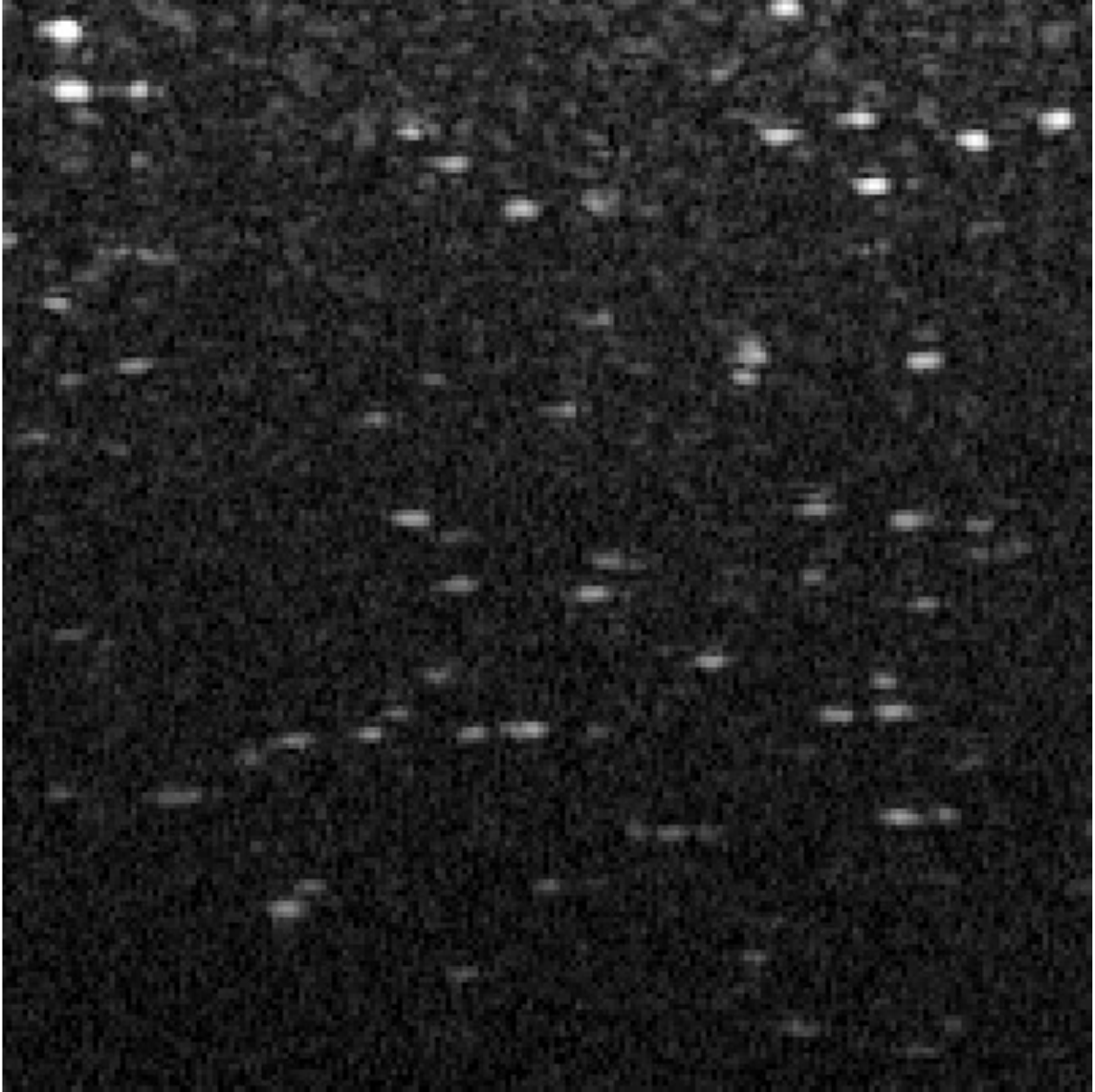}&
			\includegraphics[height=3.5cm]{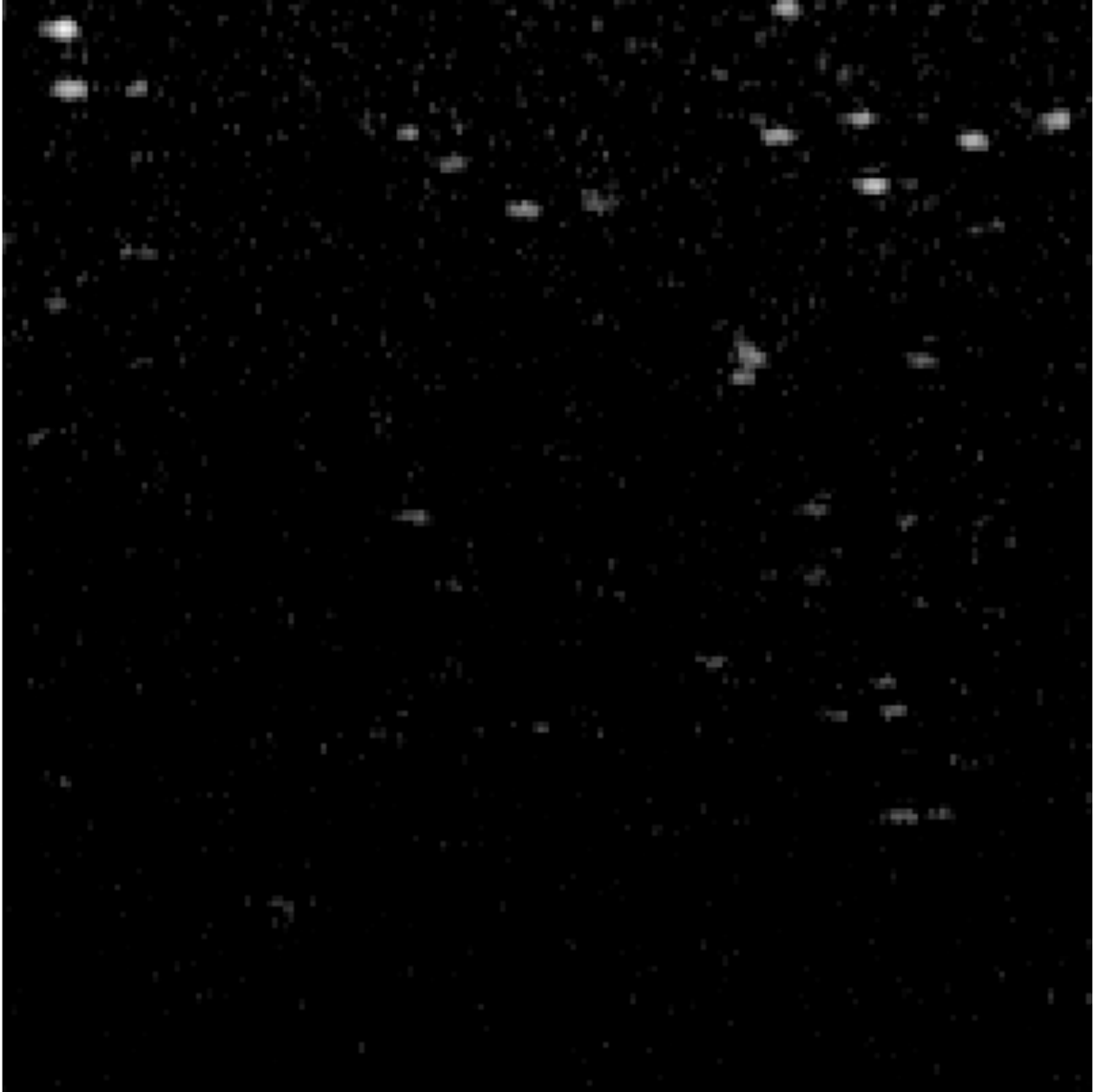}&
			\includegraphics[height=3.5cm]{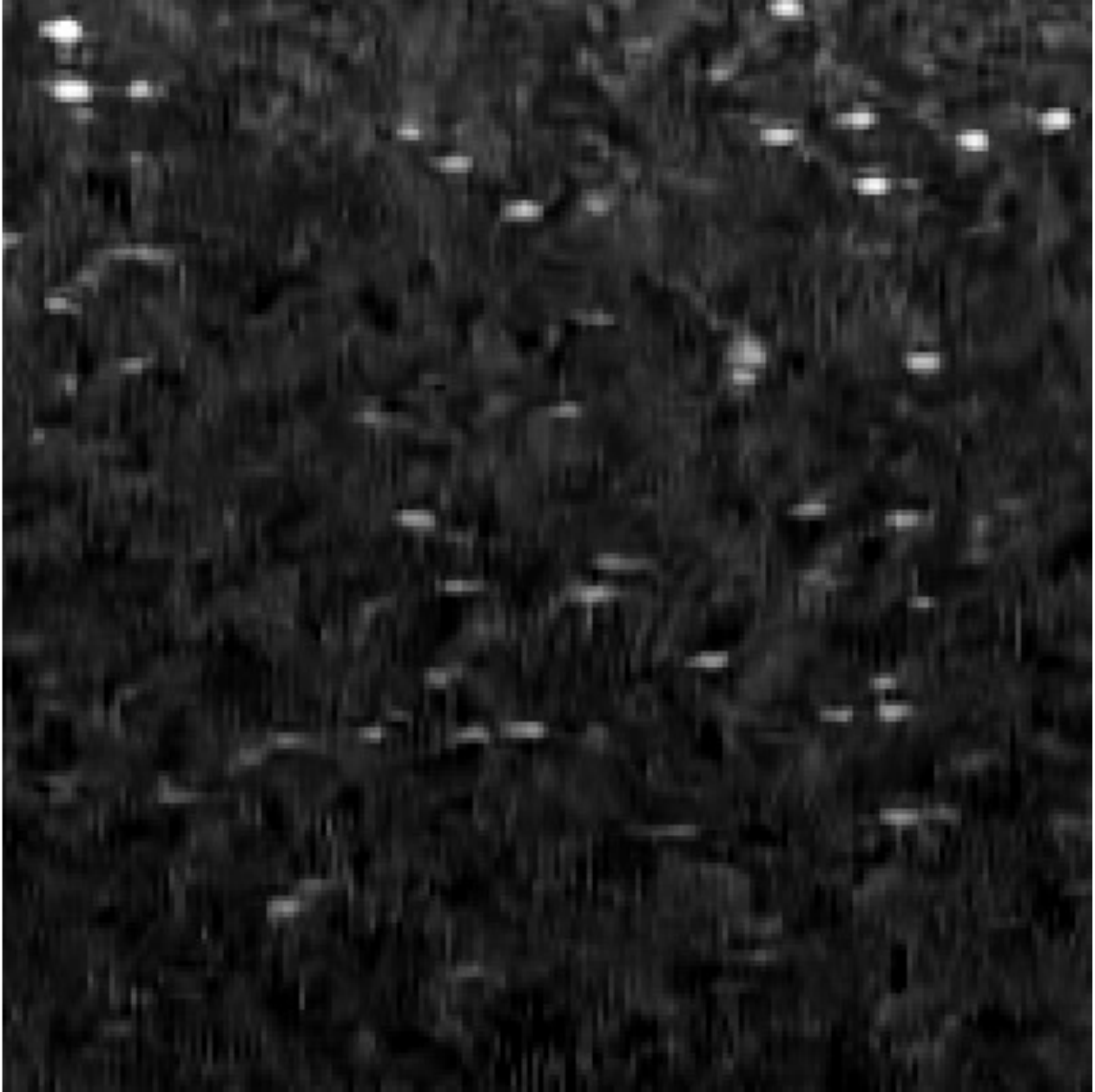}\\
			\hline
		\end{tabular}
	\end{center}
	\caption[example] 
	{\label{fig:latex_roi} Results within ROIs shown in Figure~\ref{fig:sample_latex}, showing denoising effect.}
\end{figure}

The regions from Figure~\ref{fig:sample_latex} are shown in Figure~\ref{fig:latex_roi} for various methods. These are: a single temporal frame; the B-scan average of 20 frames; the average of the 5 temporal frames after linear warping; and the result of our proposed motion compensated denoising using BM4D.

From Figure~\ref{fig:sample_latex}, the effect of our approach can be visualized. The single frame exhibits a significant amount of noise compared to the 20 frame average, especially in ROI 2. Simply applying B-scan averaging to warped frames results in a very poor image, where the brightest features are preserved, but most speckle structure is lost. In the case of the denoised images, ROI 1 appears very similar to the ground truth, with good preservation of speckle structure. In ROI 2, the micro-beads are well preserved, and the image is more clear than the single frame, albeit with some block-like artifacts.

\begin{figure}[htb!]
	\begin{center}
		\begin{tabular}{c|c|c|c|c|}
			& displacement 1 & displacement 2 & displacement 3 & displacement 4 \\
			\hline
			\rotatebox{90}{initial estimate}&
			\includegraphics[height=3.5cm]{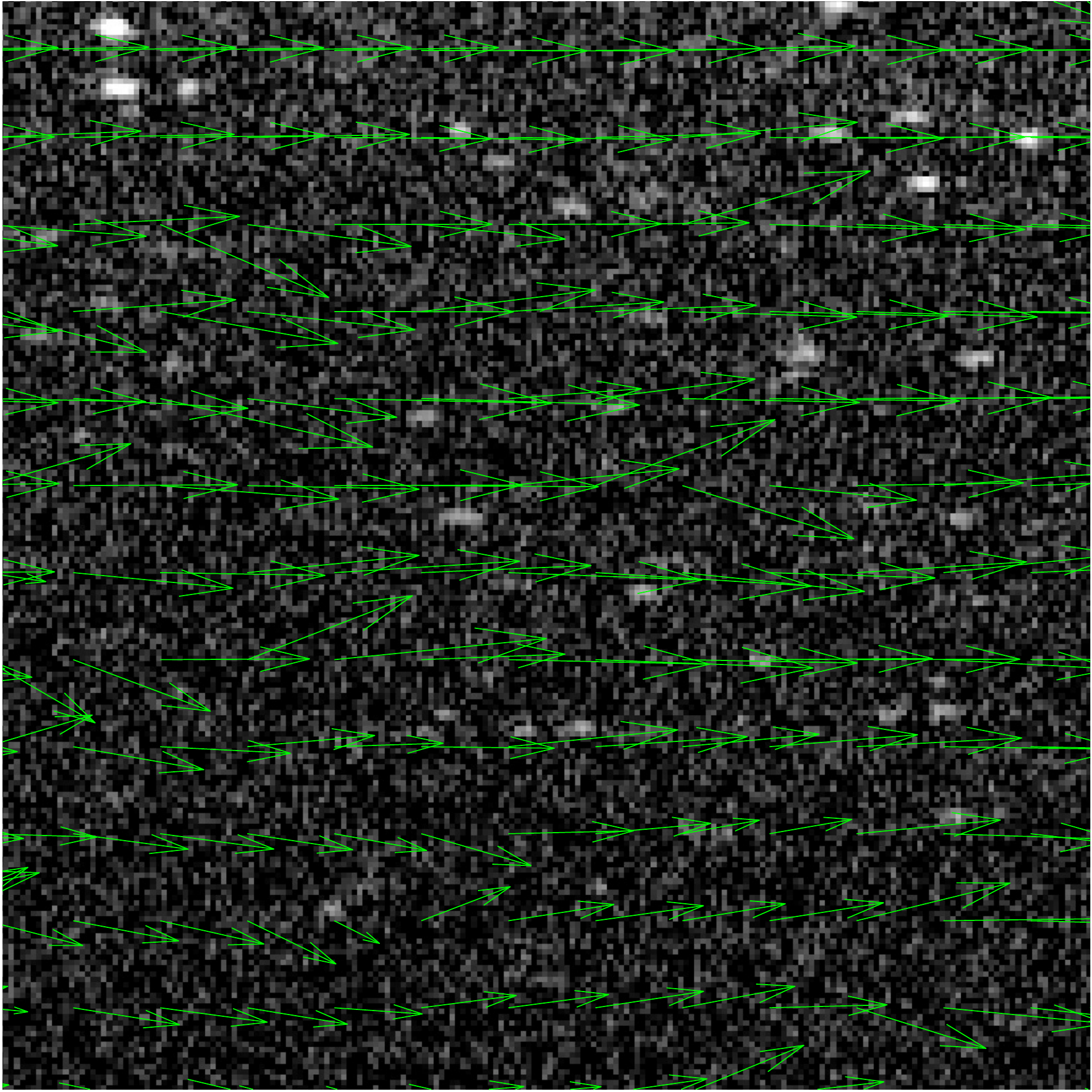}&
			\includegraphics[height=3.5cm]{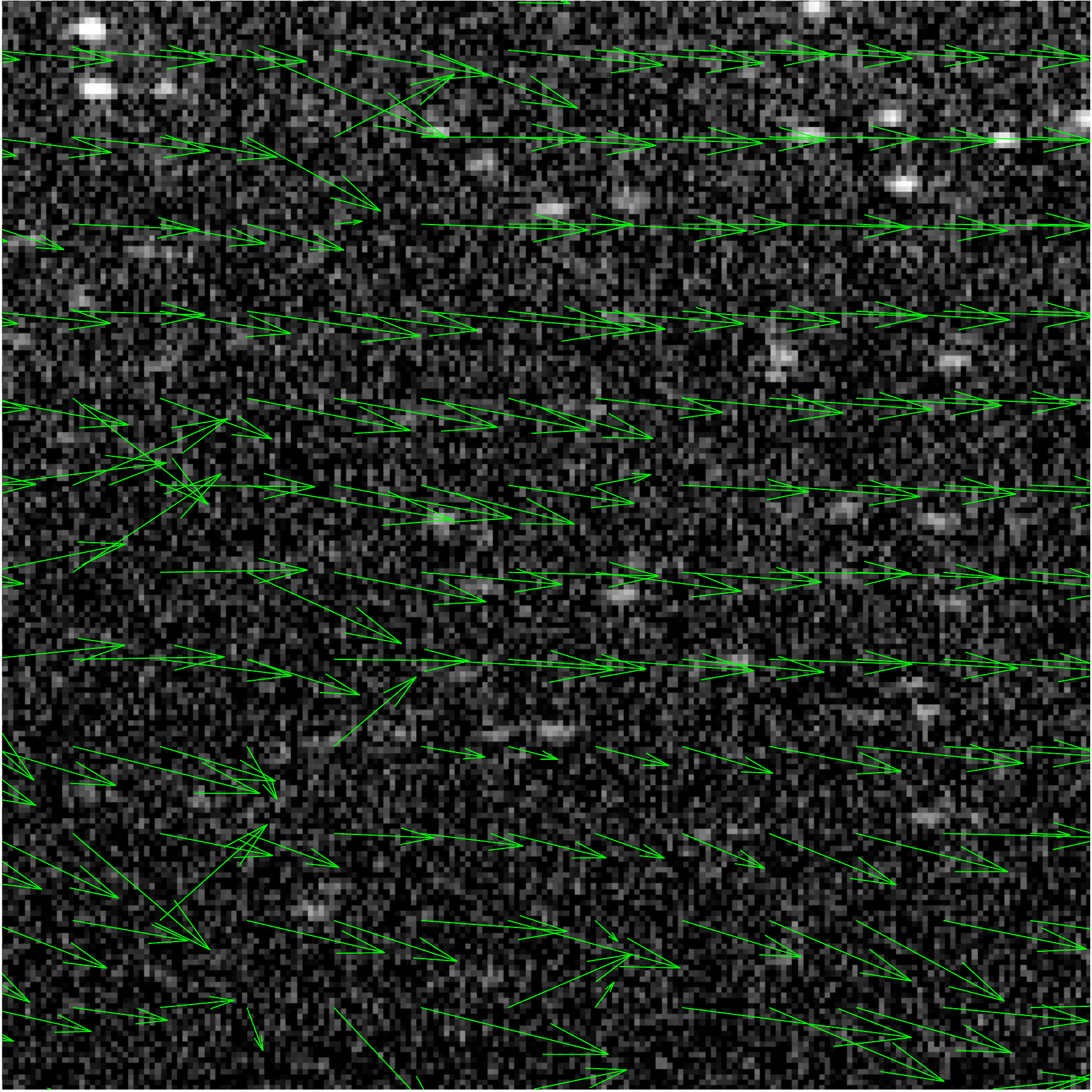}&
			\includegraphics[height=3.5cm]{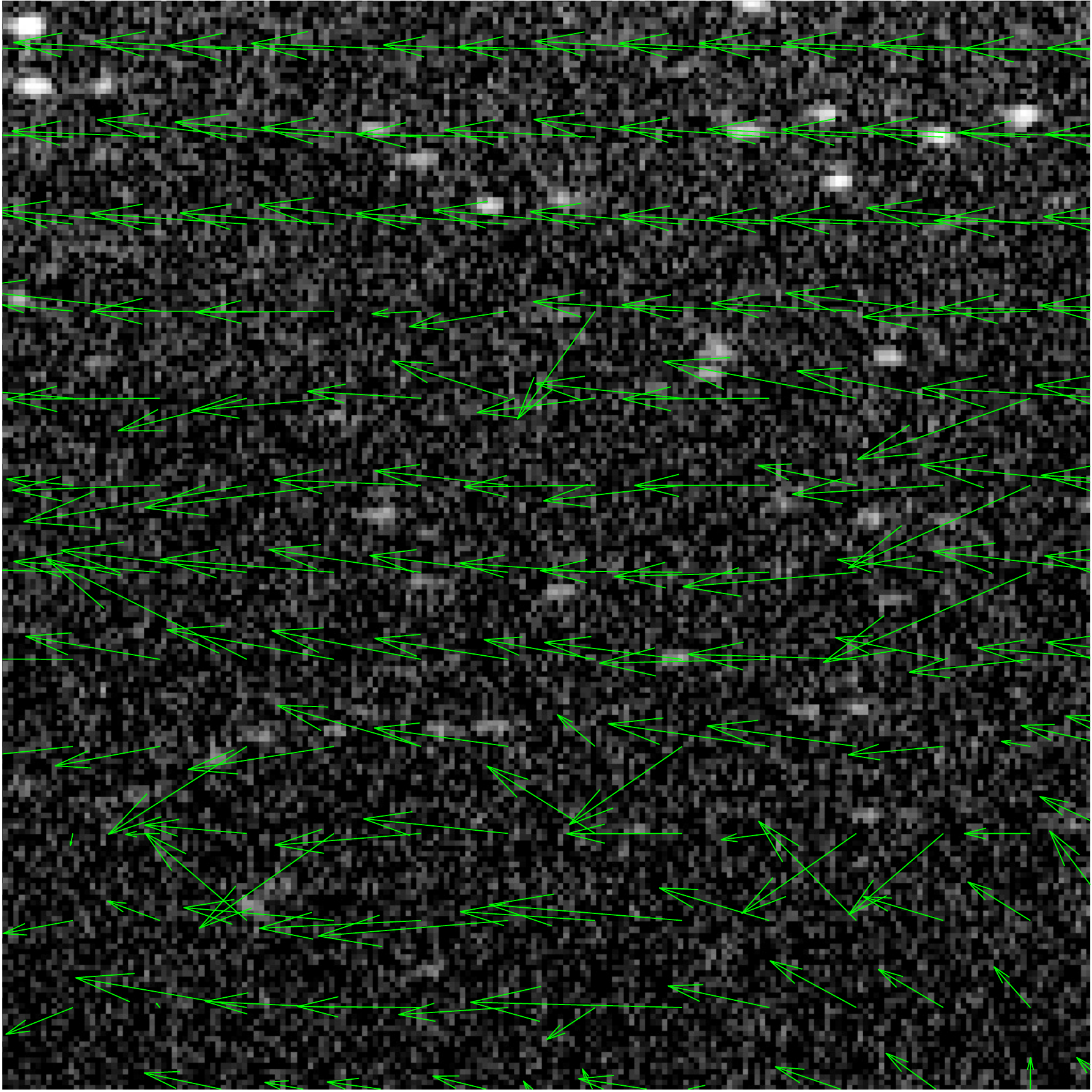}&
			\includegraphics[height=3.5cm]{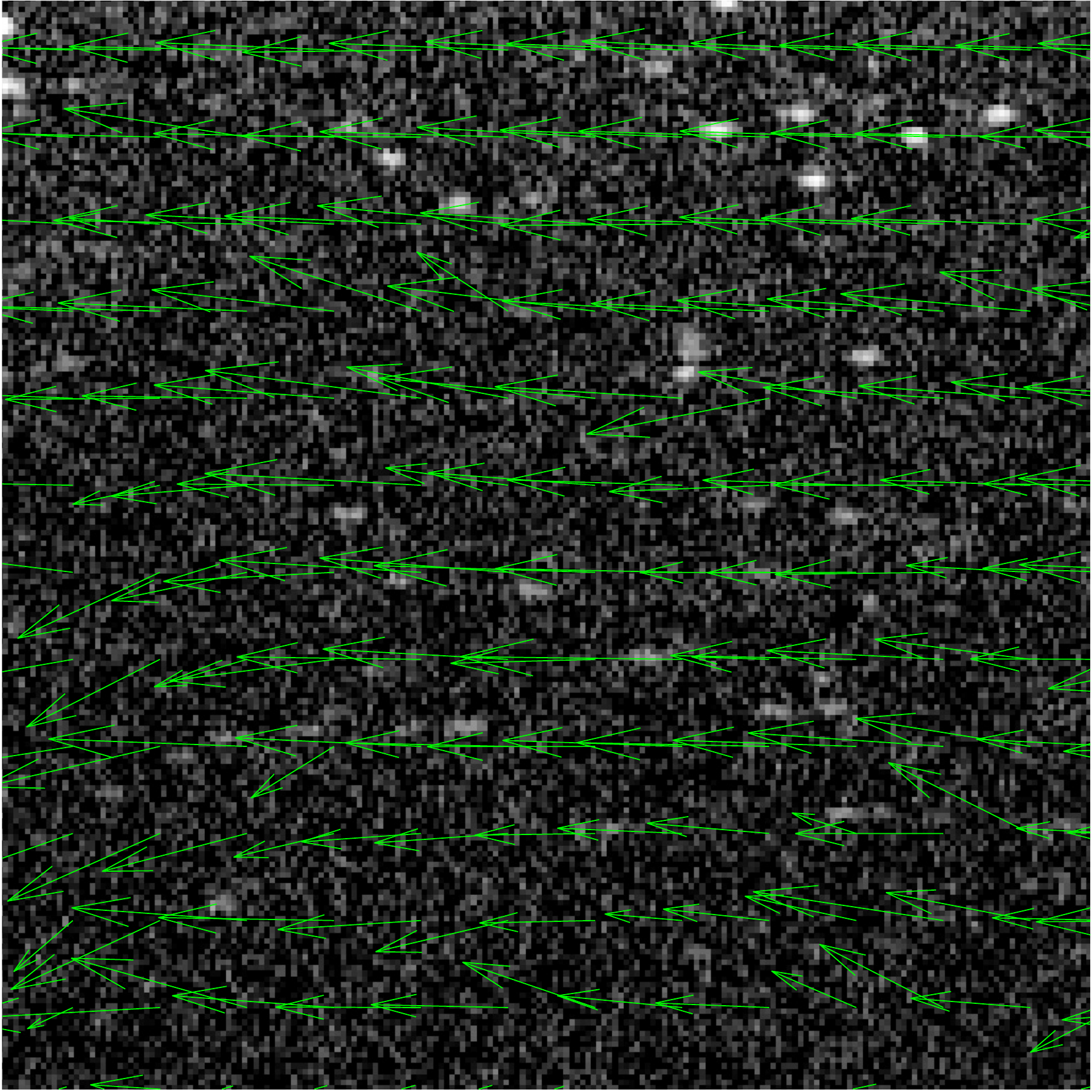}\\
			\hline
			\rotatebox{90}{denoised estimate}&
			\includegraphics[height=3.5cm]{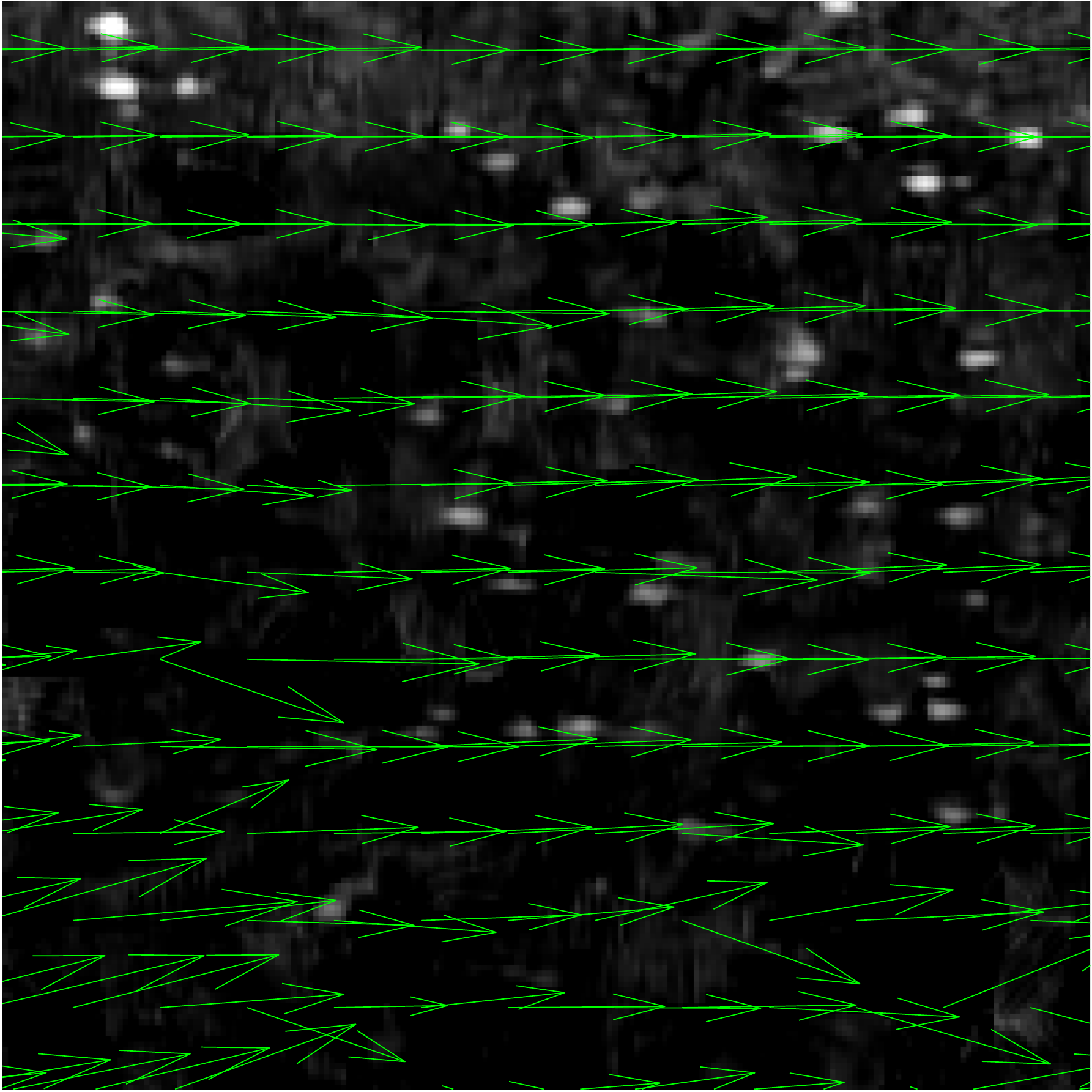}&
			\includegraphics[height=3.5cm]{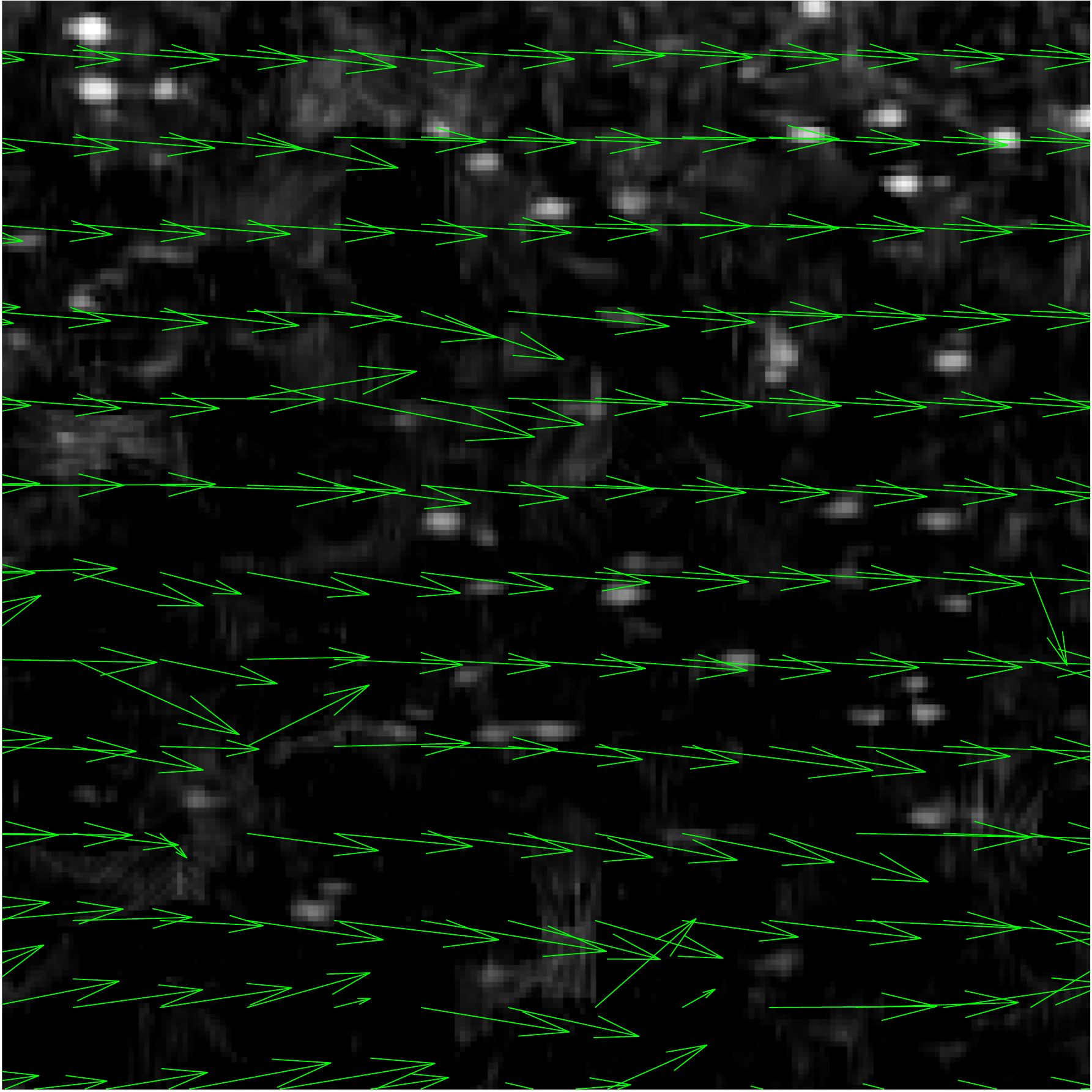}&
			\includegraphics[height=3.5cm]{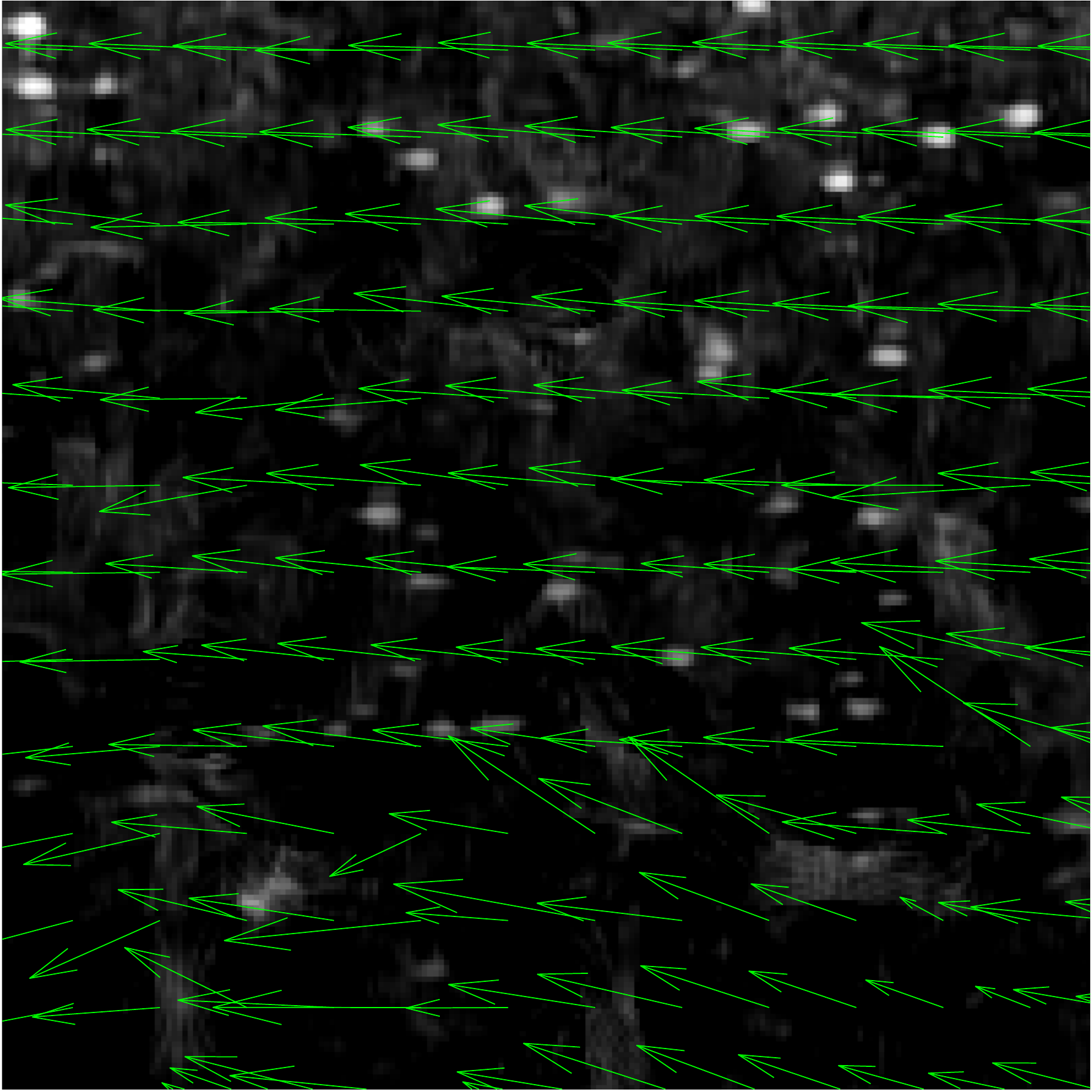}&
			\includegraphics[height=3.5cm]{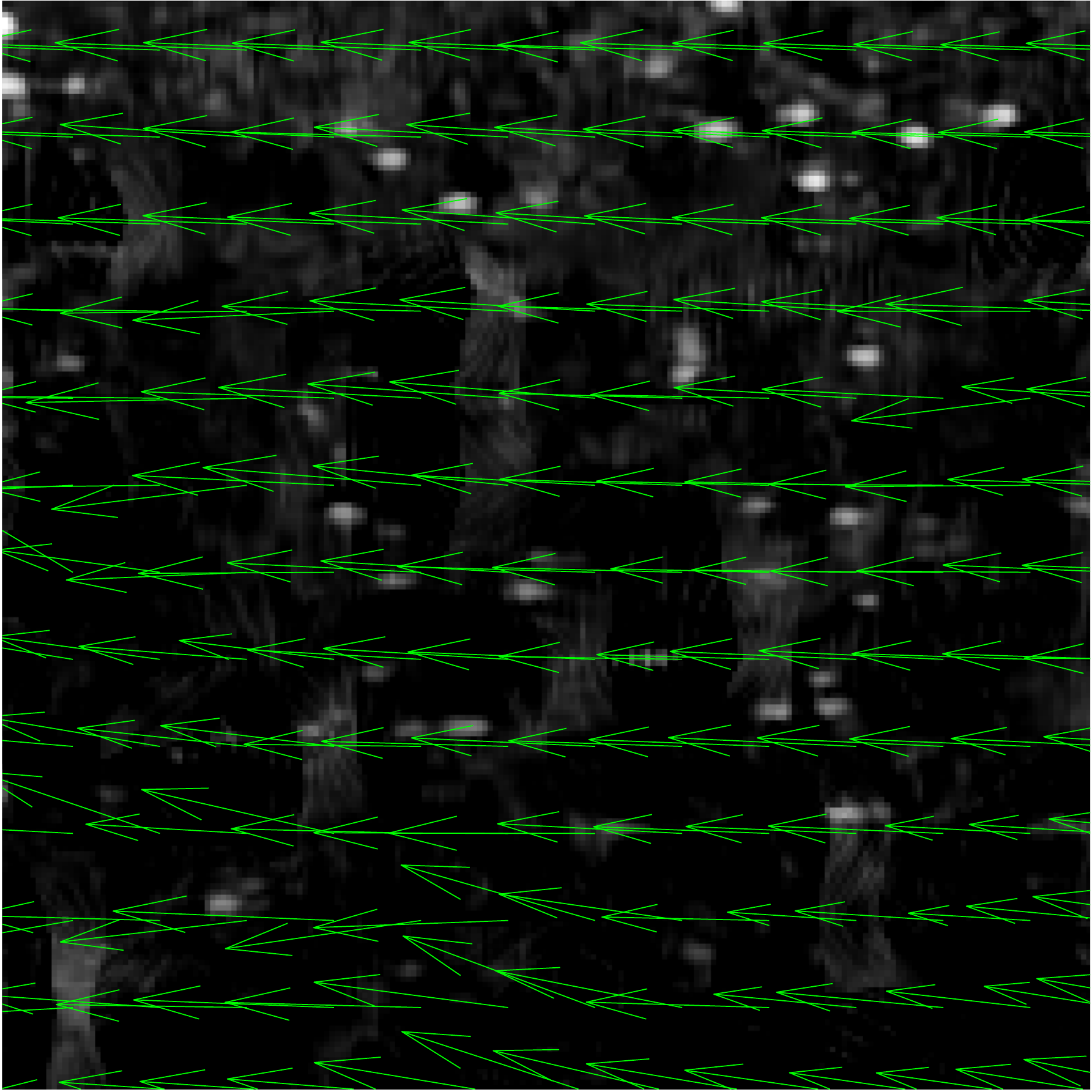}\\
			\hline
		\end{tabular}
	\end{center}
	\caption[example] 
	{\label{fig:result_latex_flow} Estimated displacements in ROI 2, before and after motion compensated denoising.}
\end{figure}

Displacements between the 5 temporal frames are shown in Figure~\ref{fig:result_latex_flow} within ROI 2, which is a challenging region due to its high level of noise. The top row shows the initial local displacement estimates from the preprocessed images, whilst the bottom row shows the re-estimated displacements after motion compensated denoising. It is clear that the uniformity of the motion estimation is dramatically enhanced in the denoised case.

\begin{table}
	\centering
	\begin{tabular}{c|c|c|c}
		method & image RMSE & x-displacement RMSE & y-displacement RMSE \\
		\hline
		original & 0.203 & 1.03 & 0.489 \\
		warped mean & 6.10 & - & - \\
		proposed framework & \textbf{0.137} & \textbf{0.909} & \textbf{0.329} 
		
	\end{tabular}
	\caption{\label{tab:results} Quantitative results from reconstructions. All results are normalized cross-correlation (NCC) as in against fully sampled ISAM image.}
\end{table}

Quantitative results are given in Table~\ref{tab:results}. Here the root--mean--squared--error (RMSE) is calculated against the ground truth image, and throughout all local displacement estimations, to give evaluation of both criteria.

Firstly, it is evident that applying a naive average throughout the warped frames, heavily degrades the image fidelity. On the other hand, BM4D throughout these frames reduces the error by 33\% against the ground truth. In terms of displacement estimation, our approach also offers a significant gain in accuracy of 12\% and 33\% in the lateral and axial dimensions respectively.

From the qualitative and quantitative results, we have shown that this proposed approach does successfully simultaneously enhance both image quality and displacement estimation.

\subsection{Complex Displacement Example}
\begin{figure}
\begin{center}
	\begin{tabular}{cc}
		\includegraphics[height=7cm]{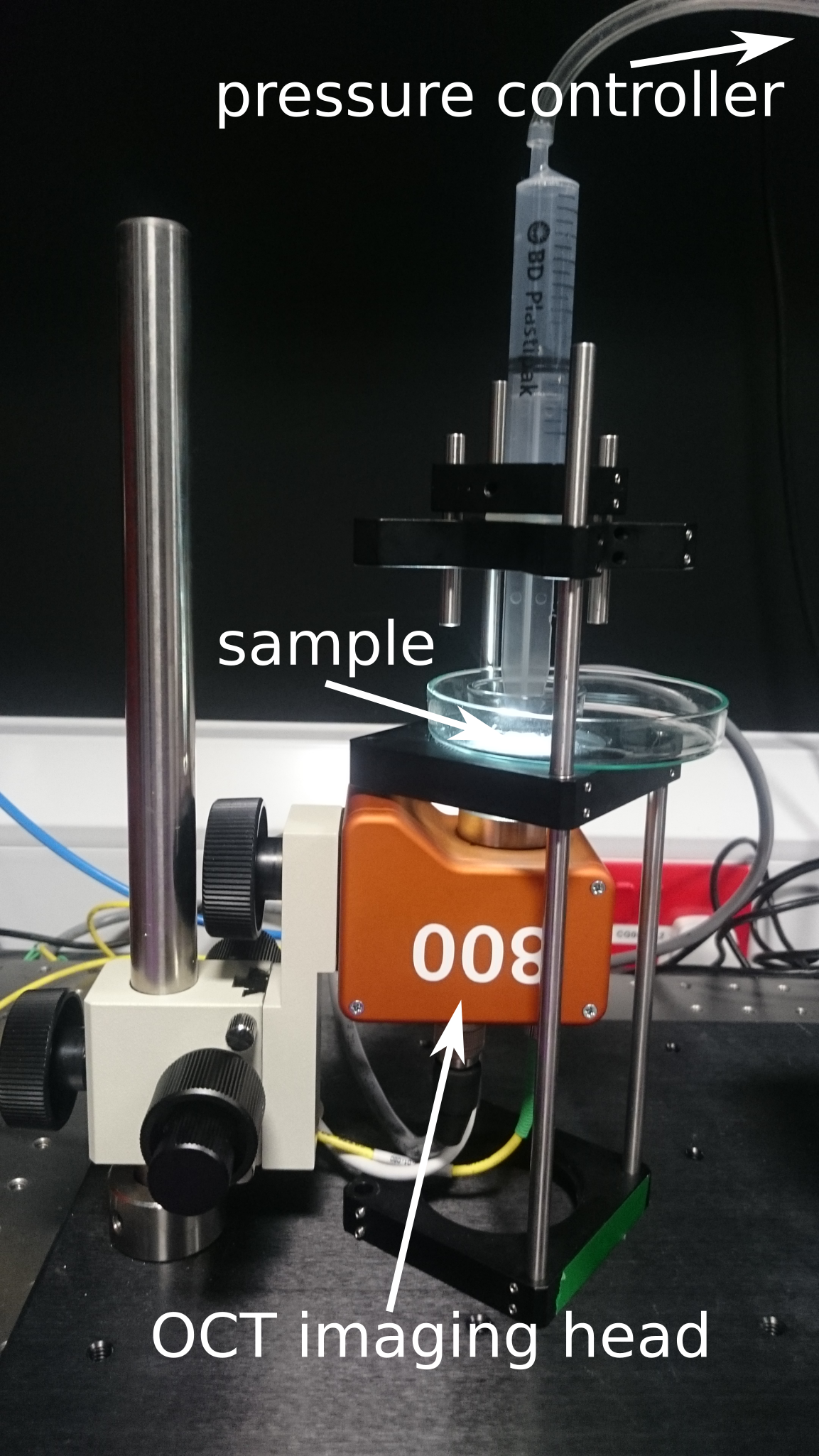}
		&\includegraphics[height=5cm]{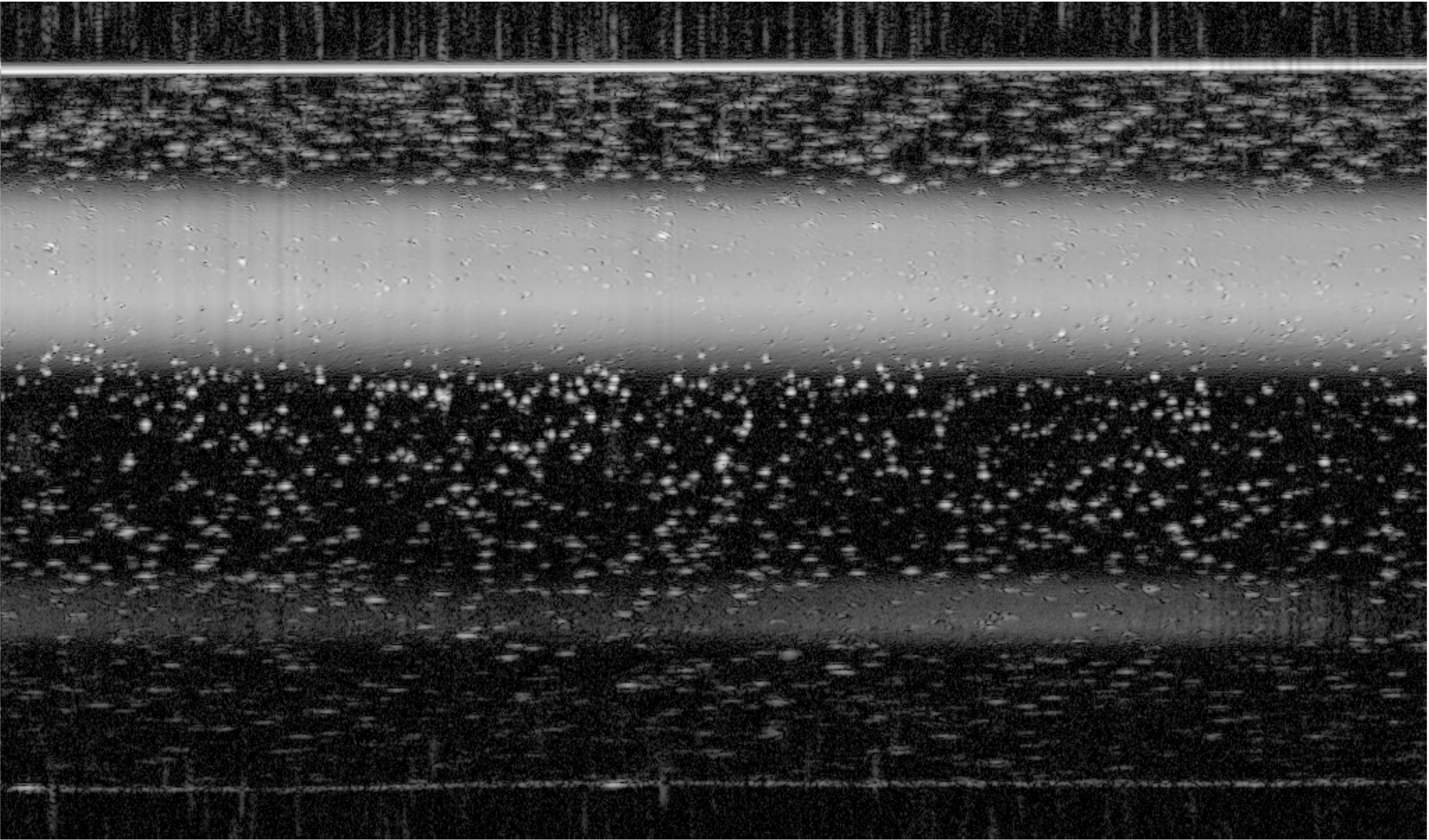}
	\end{tabular}
\end{center}
\caption[example] 
{ \label{fig:setup} 
	Setup with syringe for loading sample, along with standard IFFT reconstruction, to exhibit artifacts from the scan.}
\end{figure}

In the second case, we wish to evaluate the displacement from a more complex motion, which would be applicable to elastography. In this case, we apply a uniaxial stress on a sample of PDMS resin, by means of a syringe attached to a pressure controller --- shown in Figure~\ref{fig:setup}. 

In this setup, the sample is placed in a glass Petri dish, and imaged by the bottom. The first glass interface leads to a large artifact after applying IFFT --- visible in Figure~\ref{fig:setup} --- due to the only recording the real part of the interferometric signal. Due to a dispersion mismatch between the reference and sample arms, this appears as a burred fringe, and can be removed by cancelling the contribution from the glass interference\cite{Hofer2009}, located in the negative optical delay. Along with the stage example, we also apply ISAM resampling to complete the prepocessing.

\begin{figure}[htb!]
	\begin{center}
		\begin{tabular}{c|c|c|c|c|}
			& displacement 1 & displacement 2 & displacement 3 & displacement 4 \\
			\hline
			\rotatebox{90}{initial estimate}&
			\includegraphics[height=3.5cm]{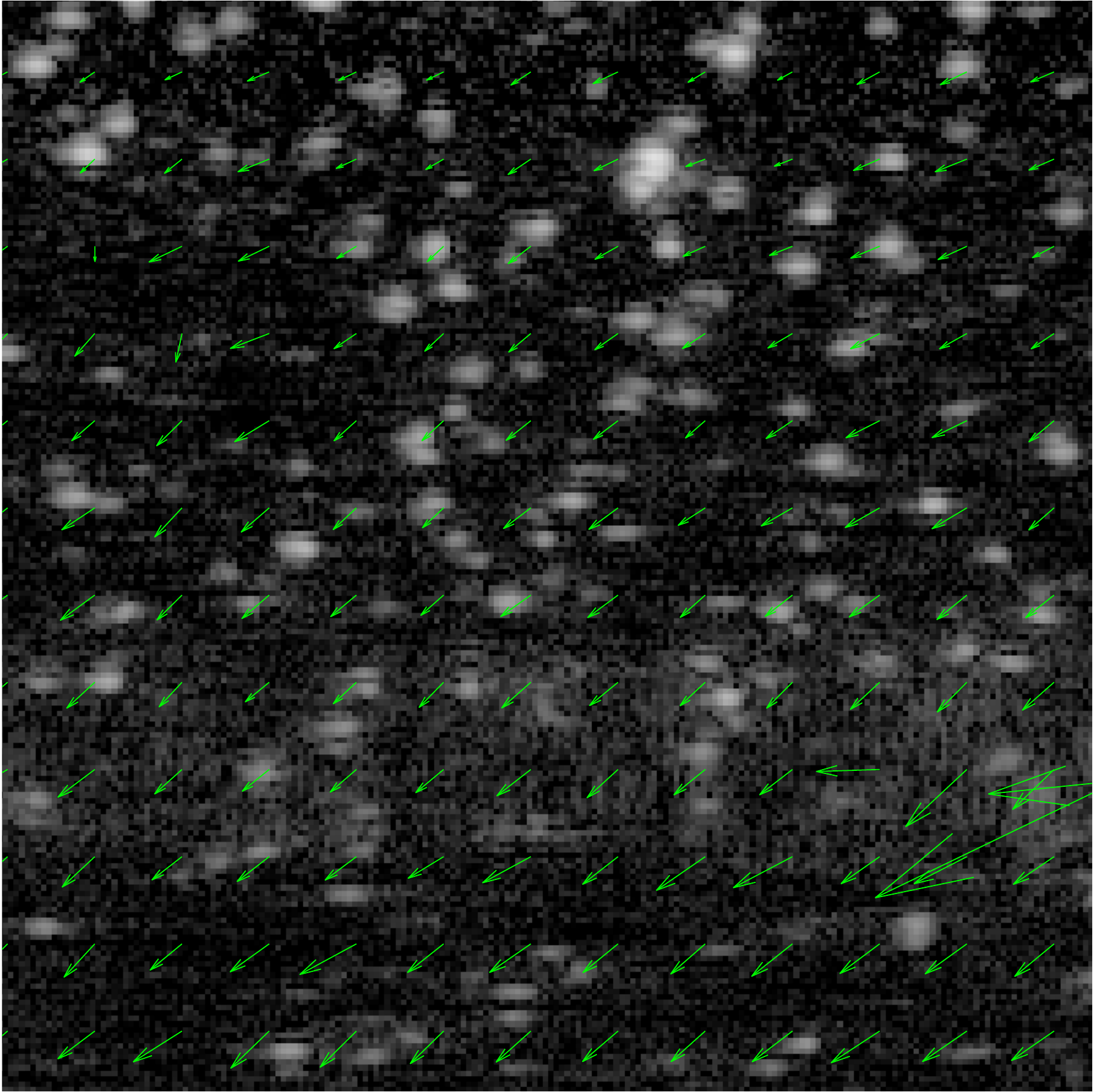}&
			\includegraphics[height=3.5cm]{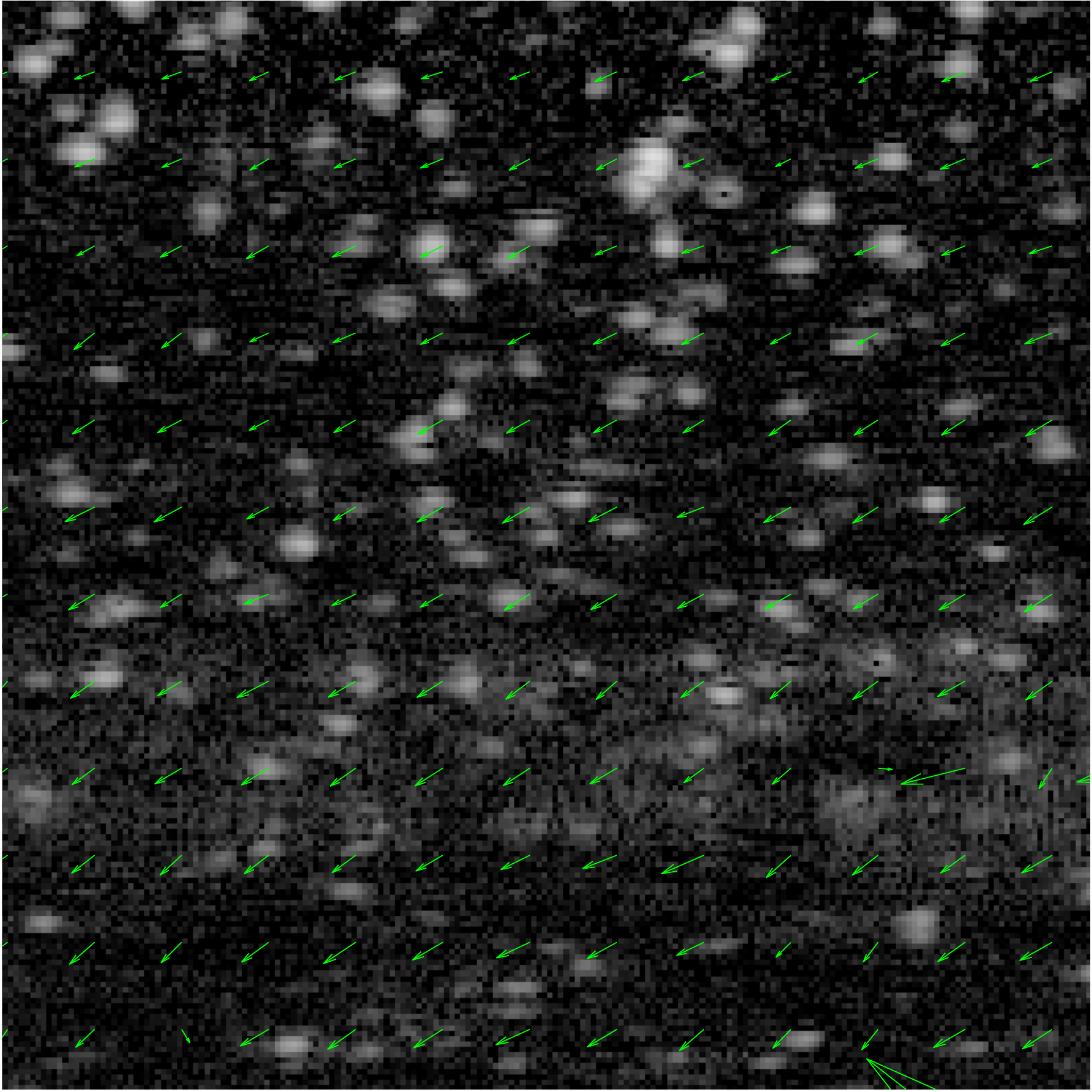}&
			\includegraphics[height=3.5cm]{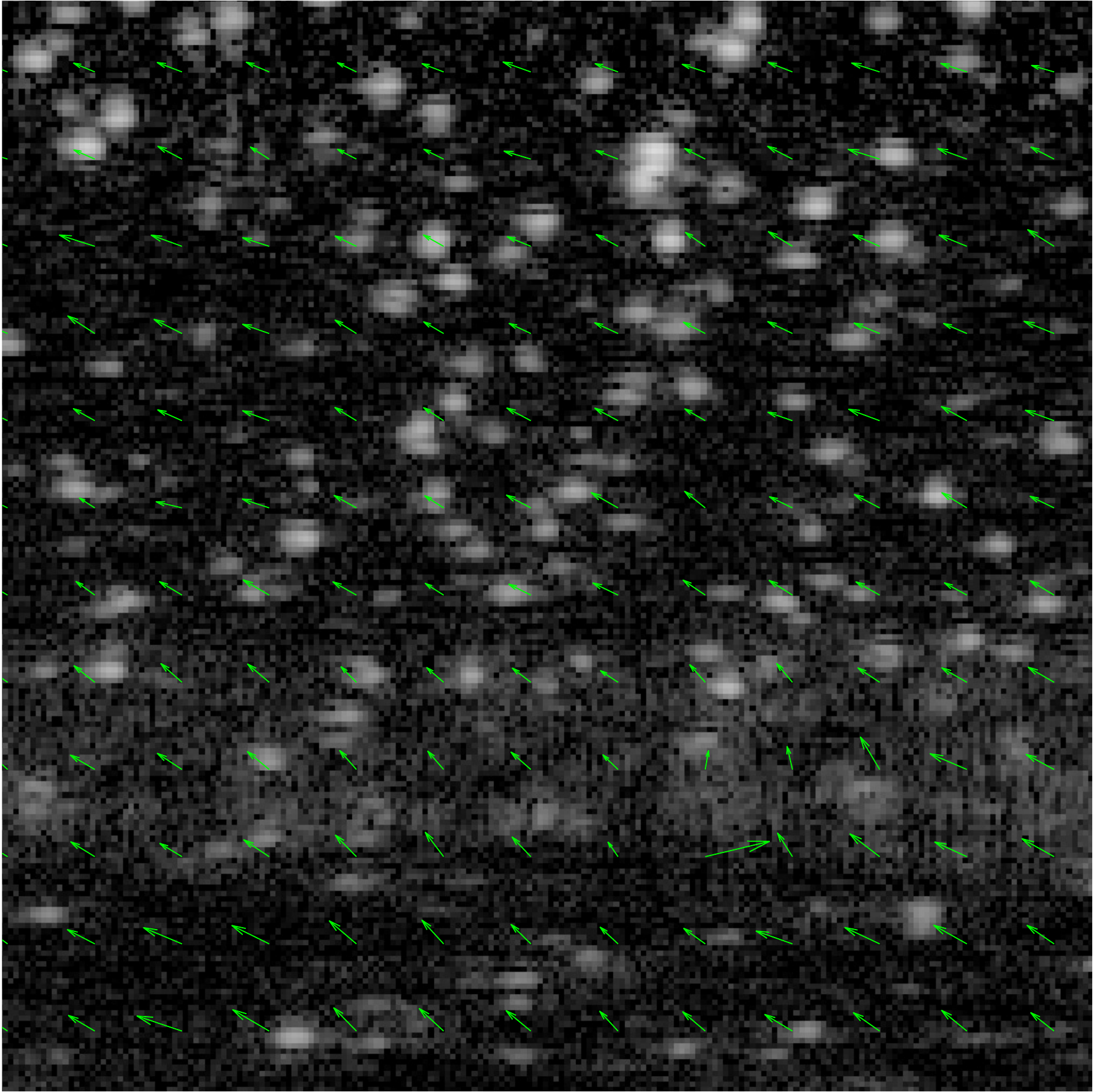}&
			\includegraphics[height=3.5cm]{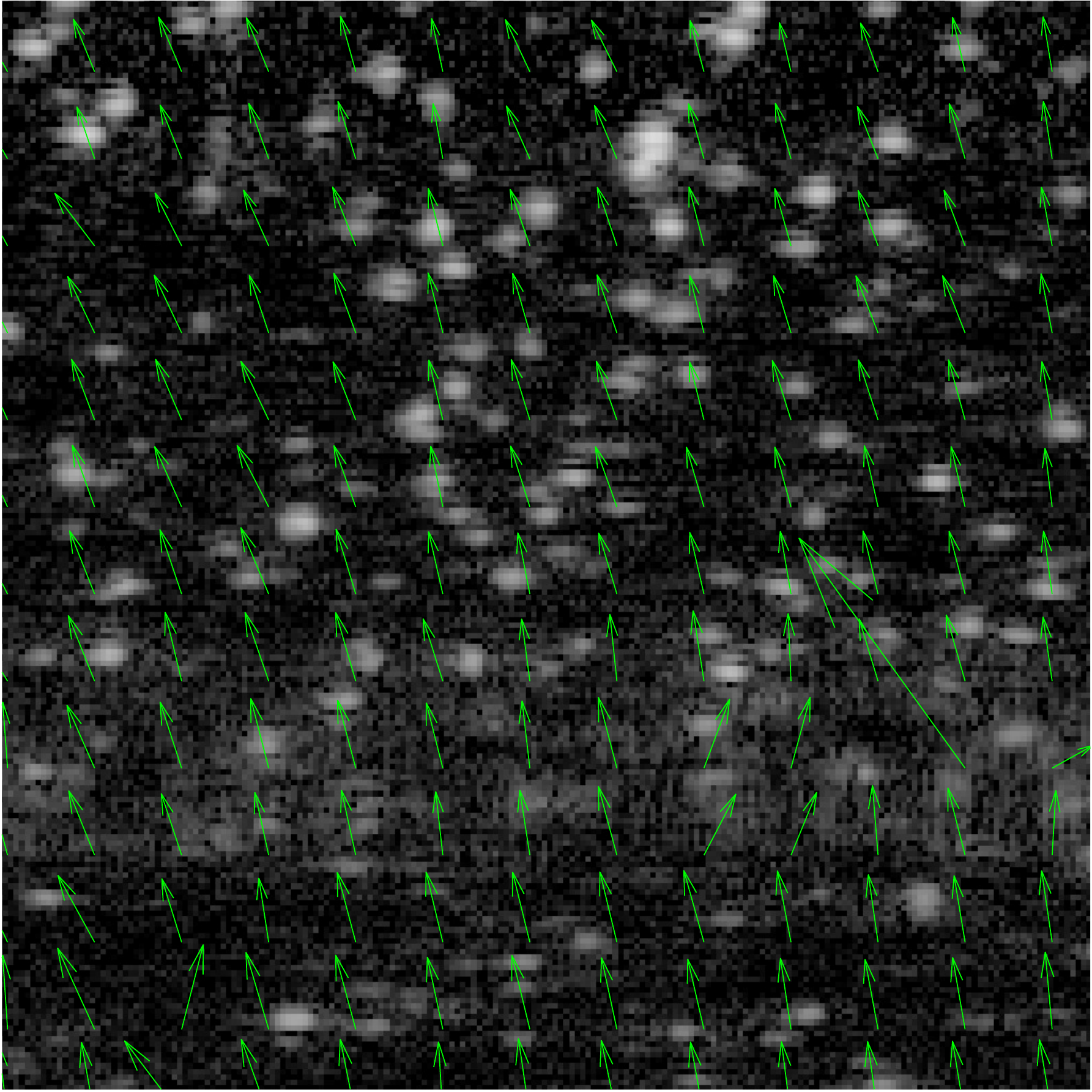}\\
			\hline
			\rotatebox{90}{denoised estimate}&
			\includegraphics[height=3.5cm]{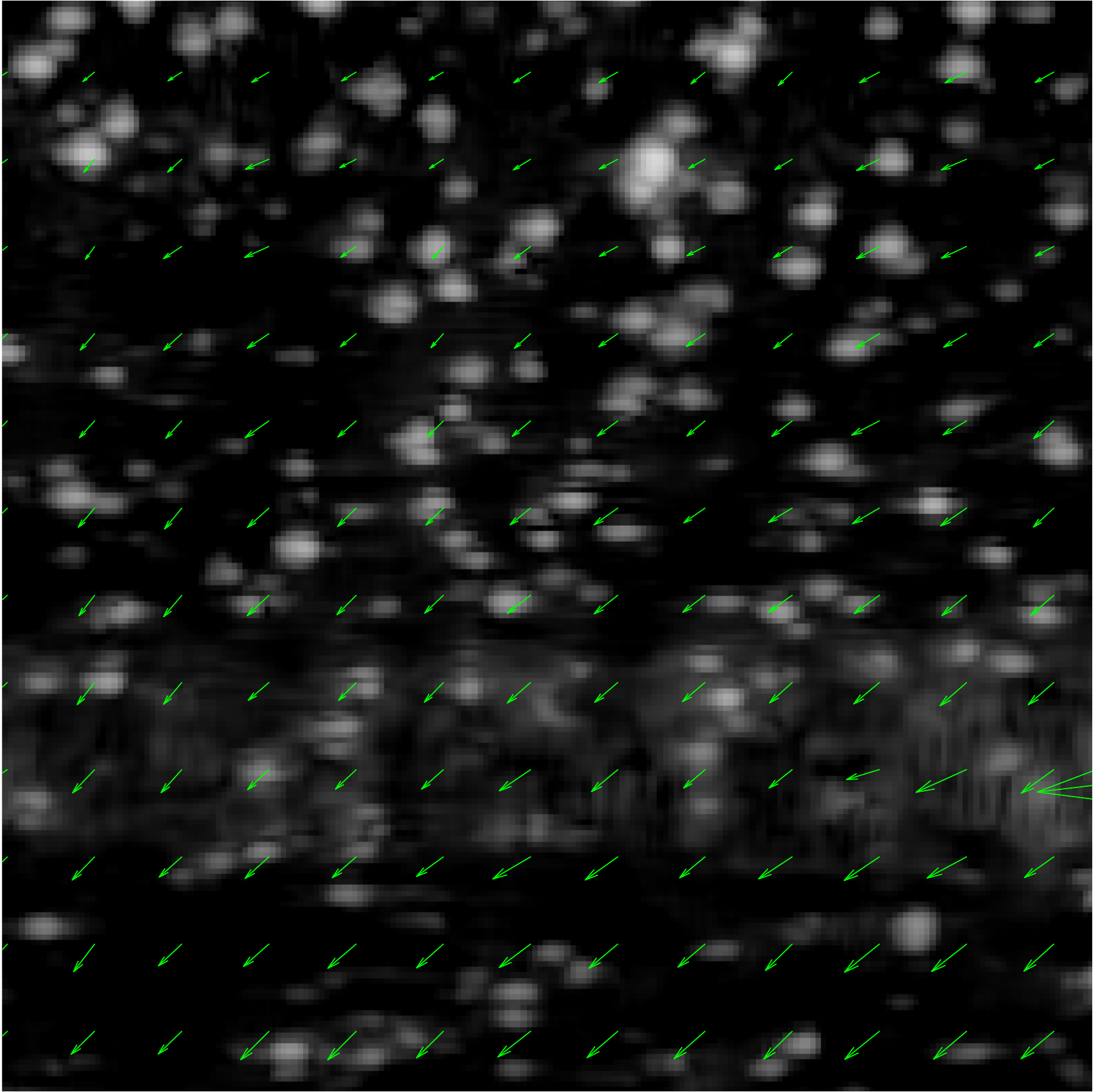}&
			\includegraphics[height=3.5cm]{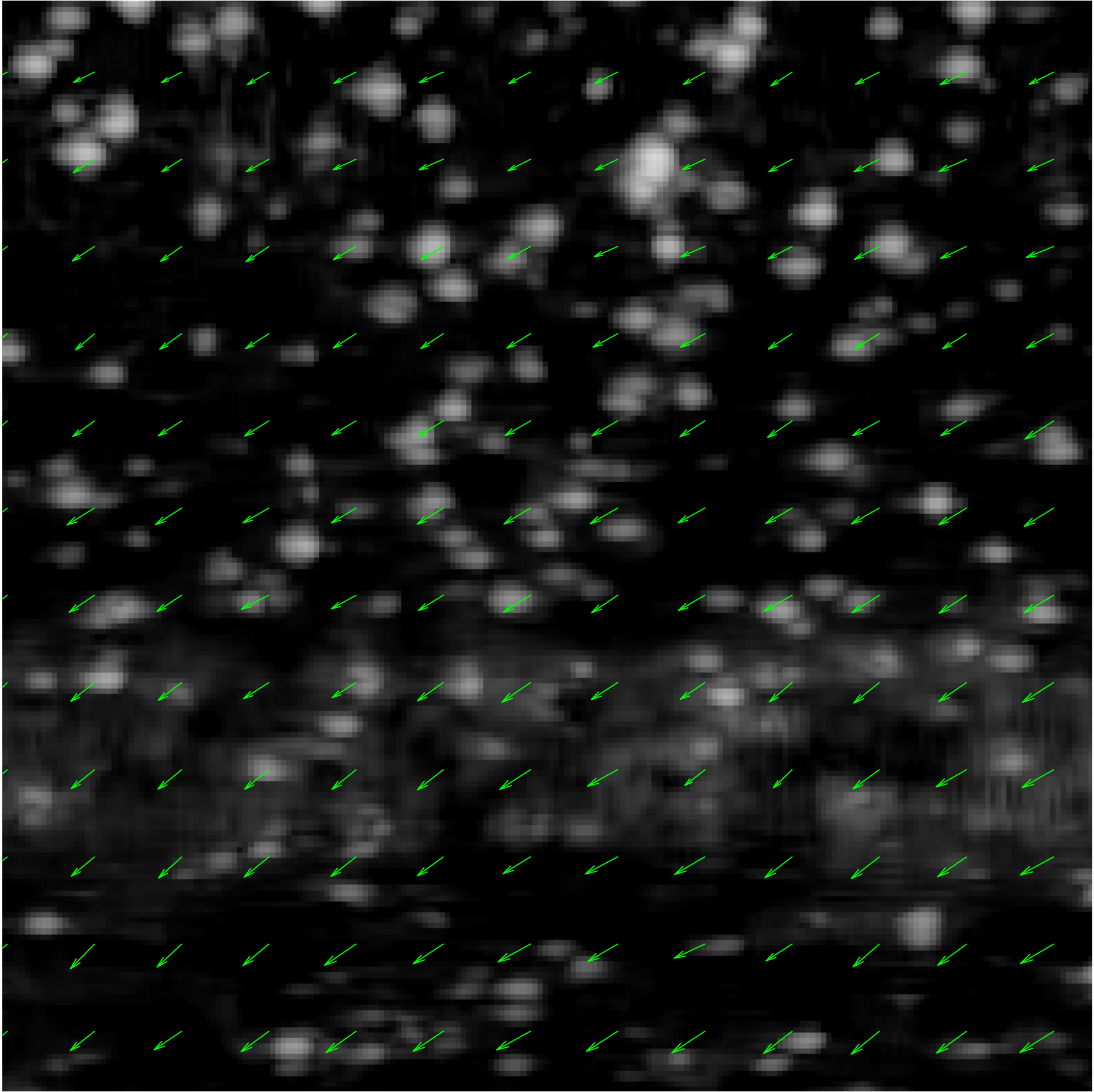}&
			\includegraphics[height=3.5cm]{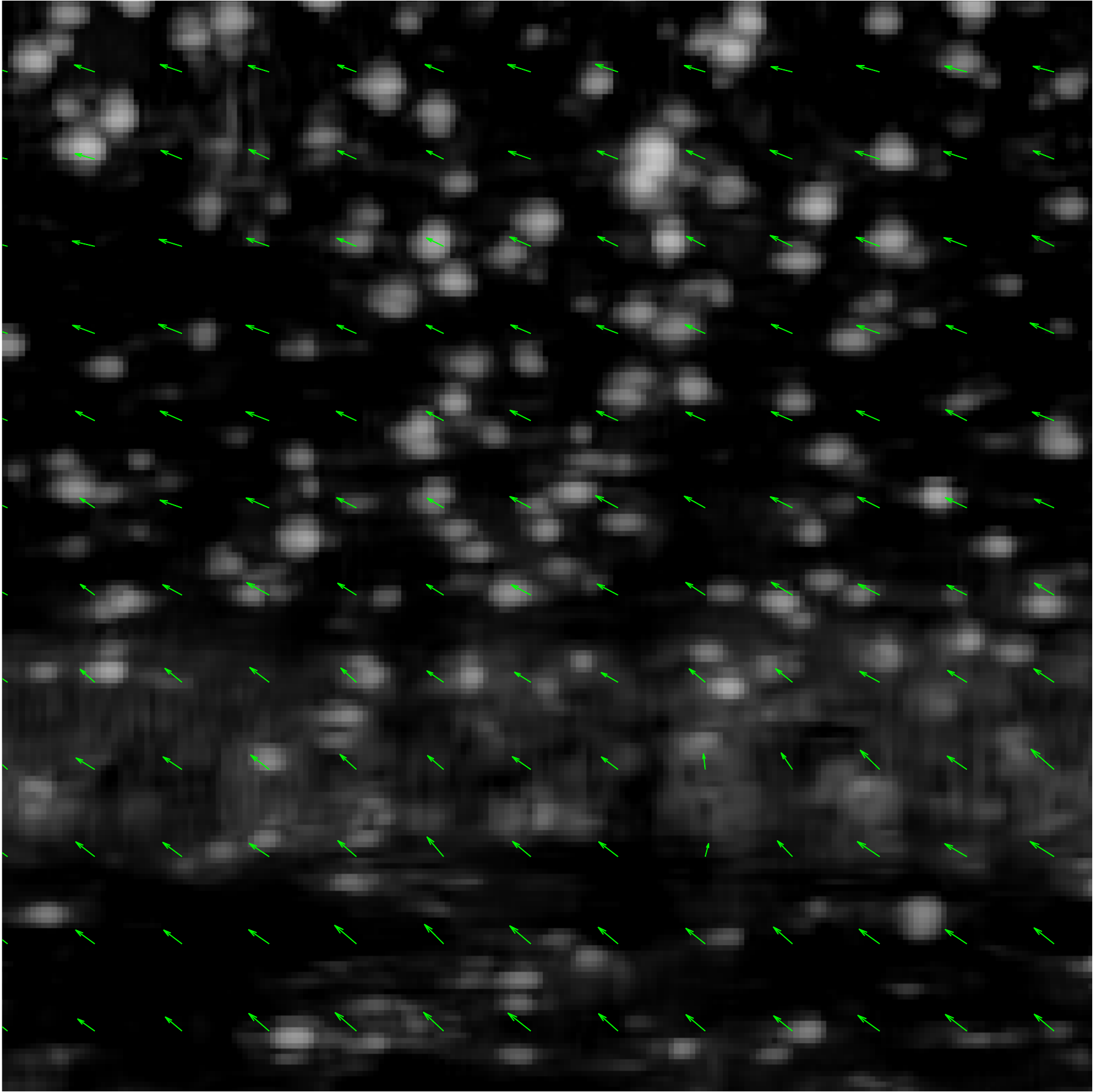}&
			\includegraphics[height=3.5cm]{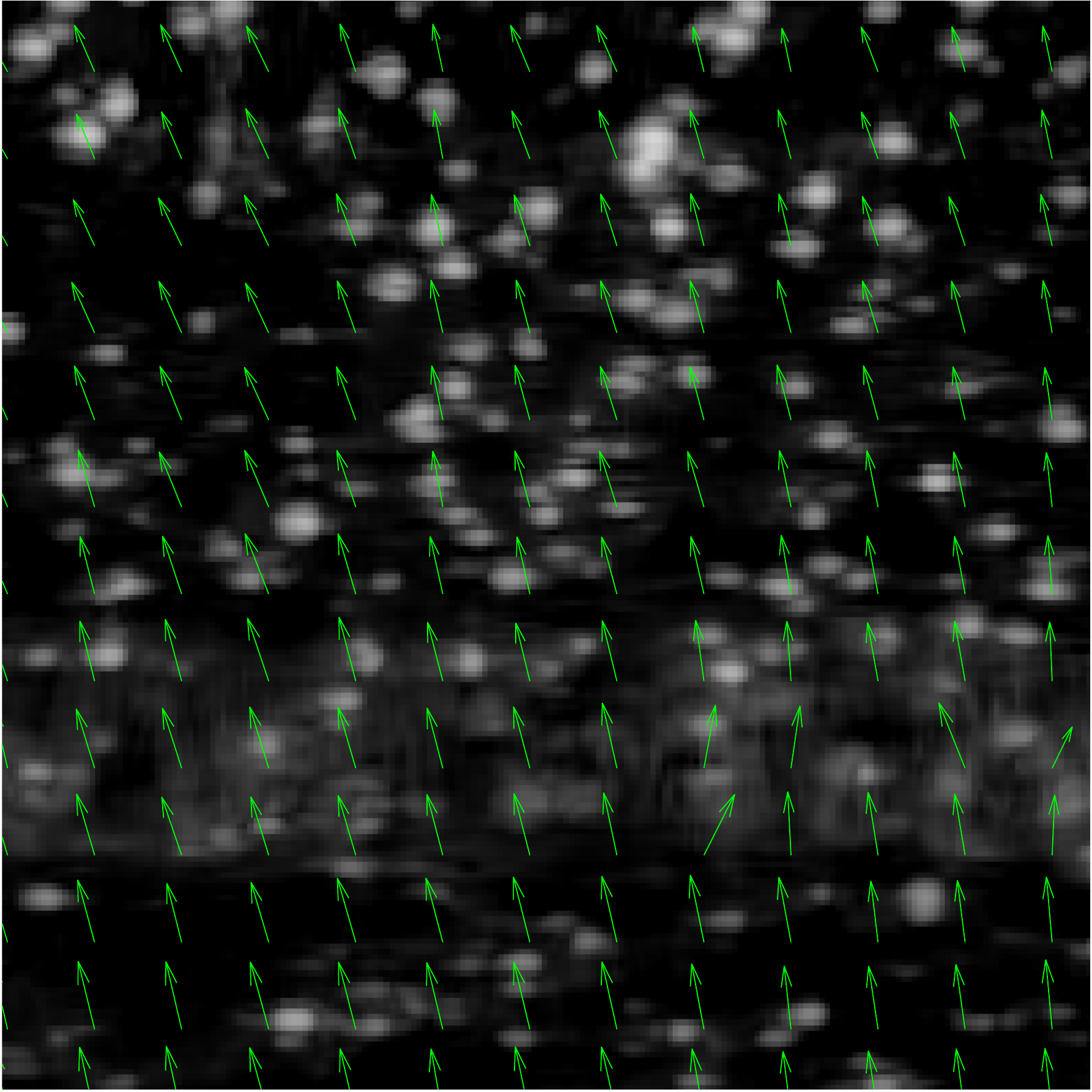}\\
			\hline
		\end{tabular}
	\end{center}
	\caption[example] 
	{\label{fig:result_squash_flow} Local displacement estimates from loaded sample, before and after motion compensated denoising.}
\end{figure}

As with the validation experiment in Section~\ref{sec:valid}, we record OCT data from 5 temporal positions, in which we apply differing force on the syringe. Ideally, this will induce a smooth non-linear deformation map throughout the sample, as it is elastically loaded. The displacement estimates from the first run, and after motion aware denoising are shown in Figure~\ref{fig:result_squash_flow}.


There are a couple of observations that can be made from the Figure~\ref{fig:result_squash_flow}. Firstly, the level of noise in the lower column is significantly reduced, whilst the structure of the scatterers is well preserved. Secondly, although there are several spurious displacement estimations in the top row, these are on the whole corrected in the lower one, which has a much more smooth deformation field with the same underlying pattern.

\section{CONCLUSIONS}
We have proposed a framework for simultaneously enhancing both image quality and local displacement accuracy, and validated it against real measurements with a commercial SD-OCT system. This allows SNR improvements, when B-scan averaging is not available due to motion, and enables more robust elastography for large displacements. Future work includes extending the method to use phase information available from SD-OCT, to increase the sub-pixel accuracy, and evaluating it with mechanical testing experiments.

\section*{ACKNOWLEDGMENTS}       
The authors sincerely thank Graham Anderson from the University of Edinburgh, for assistance creating the beaded gel phantom.
This work was supported by the UK Engineering and Physical Sciences Research Council (EPSRC) MechAScan project: EP/P031250/1.

\bibliographystyle{apalike}
\bibliography{oct_simuld} 

\begin{thebibliography}{}

\bibitem[Buades et~al., 2016]{Buades2016}
Buades, A., Lisani, J., and Miladinovi{\'{c}}, M. (2016).
\newblock {Patch-Based Video Denoising With Optical Flow Estimation}.
\newblock {\em IEEE Transactions on Image Processing}, 25(6):2573--2586.

\bibitem[Fessler and Sutton, 2003]{Fessler2003a}
Fessler, J.~A. and Sutton, B.~P. (2003).
\newblock {Nonuniform fast Fourier transforms using min-max interpolation}.
\newblock {\em IEEE Transactions on Signal Processing}, 51(2):560--574.

\bibitem[Hofer et~al., 2009]{Hofer2009}
Hofer, B., Pova{\v{z}}ay, B., Hermann, B., Unterhuber, A., Matz, G., and
  Drexler, W. (2009).
\newblock {Dispersion encoded full range frequency domain optical coherence
  tomography}.
\newblock {\em Optics Express}, 17(1):7--24.

\bibitem[Kennedy et~al., 2017]{Kennedy2017}
Kennedy, B.~F., Wijesinghe, P., and Sampson, D.~D. (2017).
\newblock {The emergence of optical elastography in biomedicine}.
\newblock {\em Nature Photonics}, 11(4):215--221.

\bibitem[Maggioni et~al., 2013]{Maggioni2013}
Maggioni, M., Katkovnik, V., Egiazarian, K., and Foi, A. (2013).
\newblock {Nonlocal Transform-Domain Filter for Volumetric Data Denoising and
  Reconstruction}.
\newblock {\em IEEE Transactions on Image Processing}, 22(1):119--133.

\bibitem[Nobach and Honkanen, 2005]{Nobach2005}
Nobach, H. and Honkanen, M. (2005).
\newblock {Two-dimensional Gaussian regression for sub-pixel displacement
  estimation in particle image velocimetry or particle position estimation in
  particle tracking velocimetry}.
\newblock {\em Experiments in Fluids}, 38(4):511--515.

\bibitem[Ralston et~al., 2007]{Ralston2007}
Ralston, T.~S., Marks, D.~L., Carney, P.~S., and Boppart, S.~A. (2007).
\newblock {Interferometric synthetic aperture microscopy}.
\newblock {\em Nature Physics}, 3(2):129--134.

\bibitem[Thielicke and Stamhuis, 2014]{Thielicke2014}
Thielicke, W. and Stamhuis, E.~J. (2014).
\newblock {PIVlab – Towards User-friendly, Affordable and Accurate Digital
  Particle Image Velocimetry in MATLAB}.
\newblock {\em Journal of Open Research Software}, 2.

\end{thebibliography}
\end{document}